\documentclass[
aps,
twocolumn,
superscriptaddress,
floatfix,
longbibliography,
nofootinbib
]{revtex4-1}
\usepackage[dvipdfmx]{graphicx}
\usepackage{dcolumn}
\usepackage{bm}
\usepackage{ulem} 
\usepackage{amsmath}
\usepackage{amssymb}
\usepackage{txfonts}
\usepackage{hyperref}
\usepackage{color} 
\usepackage{xcolor}
\usepackage{listings}
\usepackage{cprotect}

\newcommand{\bvec}[1]{\mbox{\boldmath $#1$}}

\hypersetup{
  colorlinks=true,
  linkcolor=[rgb]{0.60,0.00,0.00},
  citecolor=[rgb]{0.00,0.00,0.60},
  urlcolor=[rgb]{0.00,0.00,0.60},
  setpagesize=false
}

\begin{document}
\title{Density-Matrix Renormalization Group Study of Kitaev--Heisenberg Model on a Triangular Lattice}

\author{Kazuya Shinjo}
\affiliation{Yukawa Institute for Theoretical Physics, Kyoto University, Kyoto 606-8502, Japan}
\author{Shigetoshi Sota}
\affiliation{Computational Materials Science Research Team, RIKEN Advanced Institute for Computational Science (AICS), Kobe, Hyogo 650-0047, Japan}
\author{Seiji Yunoki}
\affiliation{Computational Materials Science Research Team, RIKEN Advanced Institute for Computational Science (AICS), Kobe, Hyogo 650-0047, Japan}
\affiliation{Computational Condensed Matter Physics Laboratory, RIKEN, Wako, Saitama 351-0198, Japan}
\affiliation{Computational Quantum Matter Research Team, RIKEN Center for Emergent Matter Science (CEMS), Wako, Saitama 351-0198, Japan}
\author{Keisuke Totsuka}
\affiliation{Yukawa Institute for Theoretical Physics, Kyoto University, Kyoto 606-8502, Japan}
\author{Takami Tohyama}
\affiliation{Department of Applied Physics, Tokyo University of Science, Tokyo 125-8585, Japan}

\date{\today}
 
\begin{abstract}
We study the Kitaev--Heisenberg model on a triangular lattice by using the two-dimensional density-matrix renormalization group method. Calculating the ground-state energy and spin structure factors, we obtain a ground-state phase diagram of the Kitaev--Heisenberg model. As suggested by previous studies, we find a 120$^\circ$ antiferromagnetic (AFM) phase, a $\mathbb{Z}_2$-vortex crystal phase, a nematic phase, a dual $\mathbb{Z}_2$-vortex crystal phase (the dual counterpart of the $\mathbb{Z}_2$-vortex crystal phase), a $\mathbb{Z}_6$ ferromagnetic phase, and a dual ferromagnetic phase (the dual counterpart of the $\mathbb{Z}_6 $ ferromagnetic phase). Spin correlations discontinuously change at phase boundaries because of first-order phase transitions. We also study the relation among the von Neumann entanglement entropy, entanglement spectrum, and phase transitions of the model. We find that the Schmidt gap closes at phase boundaries and thus the entanglement entropy clearly changes as well. This is different from the Kitaev--Heisenberg model on a honeycomb lattice, where the Schmidt gap and entanglement entropy are not necessarily a good measure of phase transitions.
\end{abstract}
\maketitle

%

\section{Introduction}
The Kitaev--Heisenberg (KH) model on a honeycomb lattice has been suggested as an effective model for (Na,Li)$_2$IrO$_3$~\cite{chaloupka2010, jiang2011, reuther2011, okamoto2013, schaffer2013, chaloupka2013, price2013, sela2014}.
The phase diagram has various exotic phases including a Kitaev spin liquid~\cite{kitaev2006}.
A strong relativistic spin-orbital interaction plays an important role in such iridate materials by entangling the spin and orbital degrees of freedom.
The $d$ orbital electrons form a $j=1/2$ Mott insulator.
A crystal field splits the $d$ orbitals into t$_{2g}$ and e$_g$ orbitals and spin-orbital interactions further yield a completely filled $j=3/2$  state and a half-filled $j=1/2$ Kramers doublet.
The exchange couplings between $j=1/2$ local moments show strong anisotropy whose easy axis strongly depends on the three distinct bond directions on the lattice.
Thus, the iridates can be effectively described by Kitaev-type interactions together with isotropic Heisenberg-type interactions~\cite{jackeli2009}.
Such interactions were originally studied by Kugel and Khomskii on the basis of the compass model~\cite{kugel1982, khaliullin2005, nussinov2015}.
To describe (Na,Li)$_2$IrO$_3$ more realistically, an extended version of the KH model with further-neighbor interactions~\cite{kimchi2011, singh2012, choi2012, reuther2014}, and additional anisotropic interactions~\cite{bhattacharjee2012, katukuri2014, rau2014, rau2014b, yamaji2014, sizyuk2014, kimchi2015, shinjo2015, chaloupka2015} has been studied.

In addition to the honeycomb lattice, the KH model on some other lattices has been proposed~\cite{kimchi2014}.
In particular, the KH model (also called the quantum compass-Heisenberg model) on a triangular lattice has attracted much attention from a theoretical viewpoint~\cite{rousochatzakis2012, becker2015} because of its rich phase diagram including a $\mathbb{Z}_2$-vortex crystal phase. 
In addition, from the experimental side, Ba$_3$IrTi$_2$O$_9$ has been suggested as a possible spin-liquid material with a triangular-lattice structure~\cite{dey2012}.
Because the KH model on a triangular lattice has both geometrical frustration and Kitaev-type frustration that breaks the SU(2) spin symmetry, the quantum effect on the model is expected to be highly nontrivial and interesting.
The possible phases of the model have been examined by classical treatments~\cite{rousochatzakis2012} as well as quantum treatments~\cite{becker2015} including the exact-diagonalization (ED) method and the density-matrix renormalization group (DMRG) method. In spite of such efforts, the presence of geometrical spin frustration and that of magnetic ordered phases with large unit cells necessitate an unbiased numerical treatment of the triangular KH model for a system with large clusters and a systematic DMRG study for full parameter range.

Motivated by these previous studies, in this paper we examine the KH model on a triangular lattice using the DMRG method for a 12$\times$6-site lattice.
We then obtain a ground-state phase diagram of the model.
As suggested in the previous studies~\cite{rousochatzakis2012, becker2015}, we found six phases: a 120$^\circ$ AFM phase, a $\mathbb{Z}_2$-vortex crystal phase, its dual counterpart, i.e., a dual $\mathbb{Z}_2$-vortex crystal phase, a nematic phase, a $\mathbb{Z}_6$ ferromagnetic (FM) phase, and its counterpart, i.e., a dual FM phase.
By calculating the spin structure factors for each spin component, we determine the magnetic structures.
The spin structure factors change discontinuously at all phase boundaries, and thus we conclude that all the phase transitions are of first order.
We find that the spin structure factors change their dominant spin component within the $\mathbb{Z}_6$ FM phase (and the dual FM phase) across the SU(2) symmetric point, although there is no phase transition.
Furthermore, we investigate the von Neumann entanglement entropy of the model.
As in the case of the spin structure factors, the entanglement entropy also discontinuously changes at all phase boundaries.
This is again due to a first-order phase transition.

In addition to the entanglement entropy, the entanglement spectrum is now accepted to be a quantity characterizing not only various phases but also phase transitions.
In fact, by using the entanglement spectrum, the phase transition between the Kitaev spin-liquid phase and magnetically ordered phases of the extended KH model on a honeycomb lattice has been discussed~\cite{shinjo2015}. 
An entanglement spectrum containing the full set of eigenvalues of the density matrix was originally proposed as a quantity characterizing the topological order of fractional quantum Hall states through a gap structure in the spectrum~\cite{li2008}. The gap is called the entanglement gap. Since the proposal, the entanglement spectrum has been studied in various systems including fractional quantum Hall systems~\cite{li2008, regnault2009, thomale2010may, lauchli2010}, topological insulators~\cite{turner2010,fidkowski2010}, spin chains~\cite{thomale2010sep}, and the Kitaev honeycomb lattice model\cite{yao2010}.
Among the gaps in the spectrum, the Schmidt gap~\cite{dechiara2012} defined by the difference between the two largest eigenvalues of the reduced density matrix, has also been found to be useful for detecting critical points through studies on the one-dimensional Kugel--Khomskii model~\cite{lundgren2012}, spin chains~\cite{dechiara2012,lepori2013, giampaolo2013}, the two-dimensional quantum Ising model~\cite{james2013}, and the spin-1/2 XXZ and spin-1 bilinear-biquadratic models on a triangular lattice~\cite{cardoner2014}. However, it has recently been pointed out that a low-energy entanglement spectrum does not necessarily provide universal information about quantum phases~\cite{chandran2014}. Furthermore, the behavior of an entanglement spectrum in two dimensions is not yet well understood.
Therefore, it is interesting to the examine entanglement spectrum of the KH model on a triangular lattice.

We find that the Schmidt gap in the triangular KH model is much larger than the other gaps among the entanglement levels.
At phase transition points, the Schmidt gap closes. 
Therefore, the change in the entanglement structure at phase transitions is clear in the case of the KH model on a triangular lattice. 
This is again due to the fact that all of the phase transitions are of first order.
This is in contrast to the extended KH model on a honeycomb lattice, where the Schmidt gap is not necessarily a measure of the phase transition~\cite{shinjo2015}.

This paper is organized as follows.
The KH model on a triangular lattice and the DMRG method are introduced in Sect.~\ref{sec2}.
In Sect.~\ref{sec3}, we show a phase diagram of the model obtained from spin-spin correlations and the ground-state energy.
The behavior of the ground-state energy, entanglement entropy, and entanglement spectrum across phase boundaries is shown.
In addition, we show the spin structure factor of each spin component at each magnetic phase and discuss the relation among the entanglement entropy, Schmidt gap, and phase transition in the model.
Finally, a summary and outlook are given in Sect.~\ref{sec4}.

\section{Model and Method}
\label{sec2}
\begin{figure}[t]
\begin{center}
\includegraphics[width=18pc]{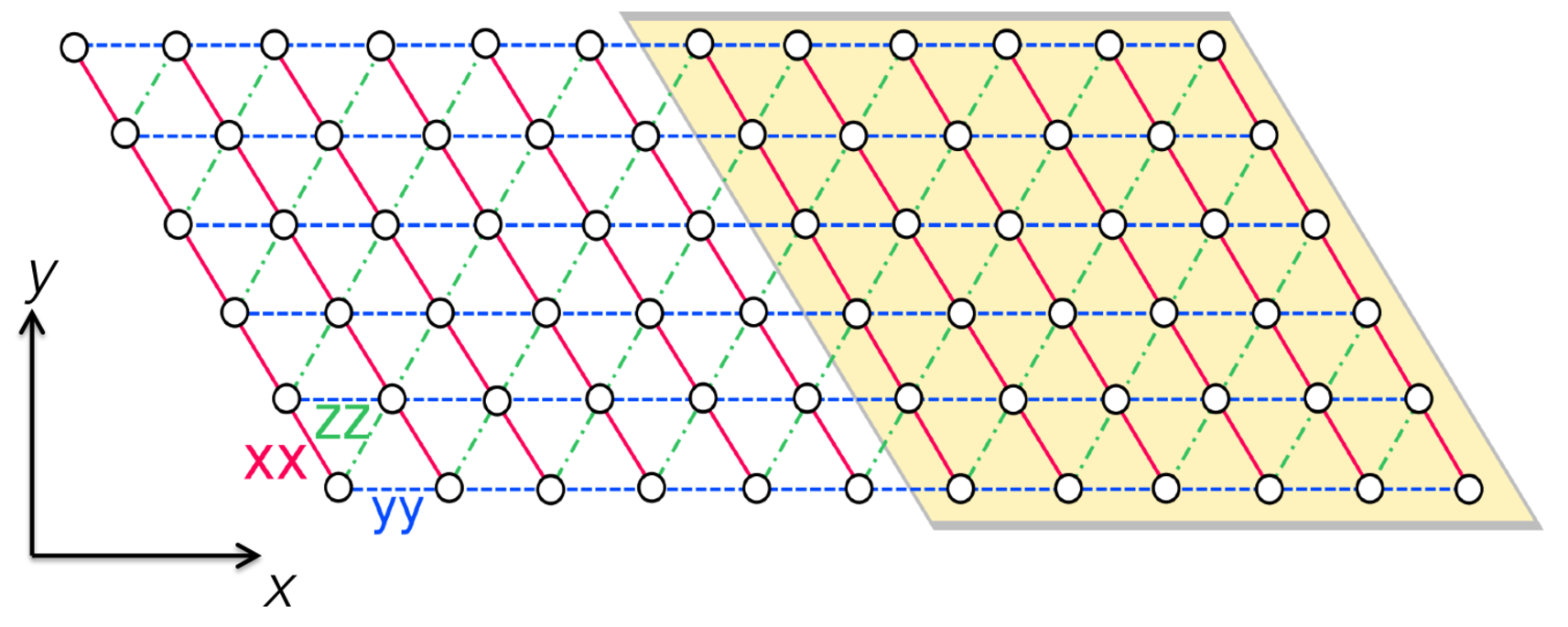}
\caption{(Color online) Triangular lattice with 12$\times$6 sites. Red solid, blue dashed, and green dashed-dotted bonds labeled by $xx$, $yy$, and $zz$ have $S^xS^x$, $S^yS^y$, and $S^zS^z$ terms in a Kitaev model, respectively. The half of the system shaded orange is the subsystem used to calculate the reduced density matrix in Sect.~\ref{sec3.3}}
\label{triangle}
\end{center}
\end{figure}

In the triangular lattice, there are three different kinds of bonds (labeled by $\gamma \gamma = xx,yy,zz$) attached to each site.
In the Kitaev term, we assign a nearest-neighbor spin-spin interaction of the form $\hat{S}_{i}^{\gamma}\hat{S}_{j}^{\gamma}$ ($\gamma=x,y,z$ is not summed over) to each species of bonds, as shown in Fig.~\ref{triangle}.
The Hamiltonian of the KH model on a given nearest-neighbor $\langle ij \rangle$ bond of type-$\gamma$ is, thus, given by 
\begin{equation}
\mathcal{H}_{\langle ij \rangle}^{(\gamma)}= \left[J_K \hat S_i^\gamma \hat S_j^\gamma + J_H \bvec{\hat S}_i \cdot  \bvec{\hat S}_j \right],
\label{KHmodel}
\end{equation}
where $\bvec{\hat S}_i$ is a spin-$1/2$ operator on site $i$, and $\gamma$ represents a combination of spin component $\gamma \in \{ x, y, z \}$ on the $xx$, $yy$, and $zz$ bonds. 
The first term is the Kitaev interaction that has $S^\gamma S^\gamma$ on the $\gamma \gamma$ bond on the triangular lattice, thereby breaking SU(2) spin rotation invariance.
The second term is the usual Heisenberg interaction with SU(2) spin symmetry.
The total Hamiltonian of the KH model is given by $H_\mathrm{KH}=\sum_{\gamma} \sum_{\langle ij \rangle \in \gamma \gamma} \mathcal{H}_{\langle ij \rangle}^{(\gamma)}$, summing over all possible nearest-neighbor pairs.

We calculate the ground state of this model by using the DMRG method~\cite{white1992, schollwock2005}.
The DMRG calculations are carried out under both cylindrical (i.e., periodic along $xx$-bond direction but open along $yy$-bond direction) and toroidal (i.e., periodic along both $xx$- and $yy$-bond directions) boundary conditions.
Unless otherwise noted, we use a system with 12 (along $yy$-bond direction) $\times$ 6 (along $xx$-bond direction) sites, i.e., a 72-site lattice as shown in Fig.~\ref{triangle}.
To perform the DMRG method, we map the original system to a snakelike one-dimensional chain along the $xx$-bond direction.
We keep 1300--1800 states in the DMRG block and perform more than 10 sweeps, resulting in a typical truncation error of $10^{-5}$ or smaller.

\section{Results and Discussions}
\label{sec3}
\subsection{Phase diagram}
\label{sec3.1}

\begin{figure}[t]
  \begin{center}
    \begin{tabular}{r}
      \begin{minipage}{1\hsize}
        \begin{center}
          \includegraphics[clip, width=20pc]{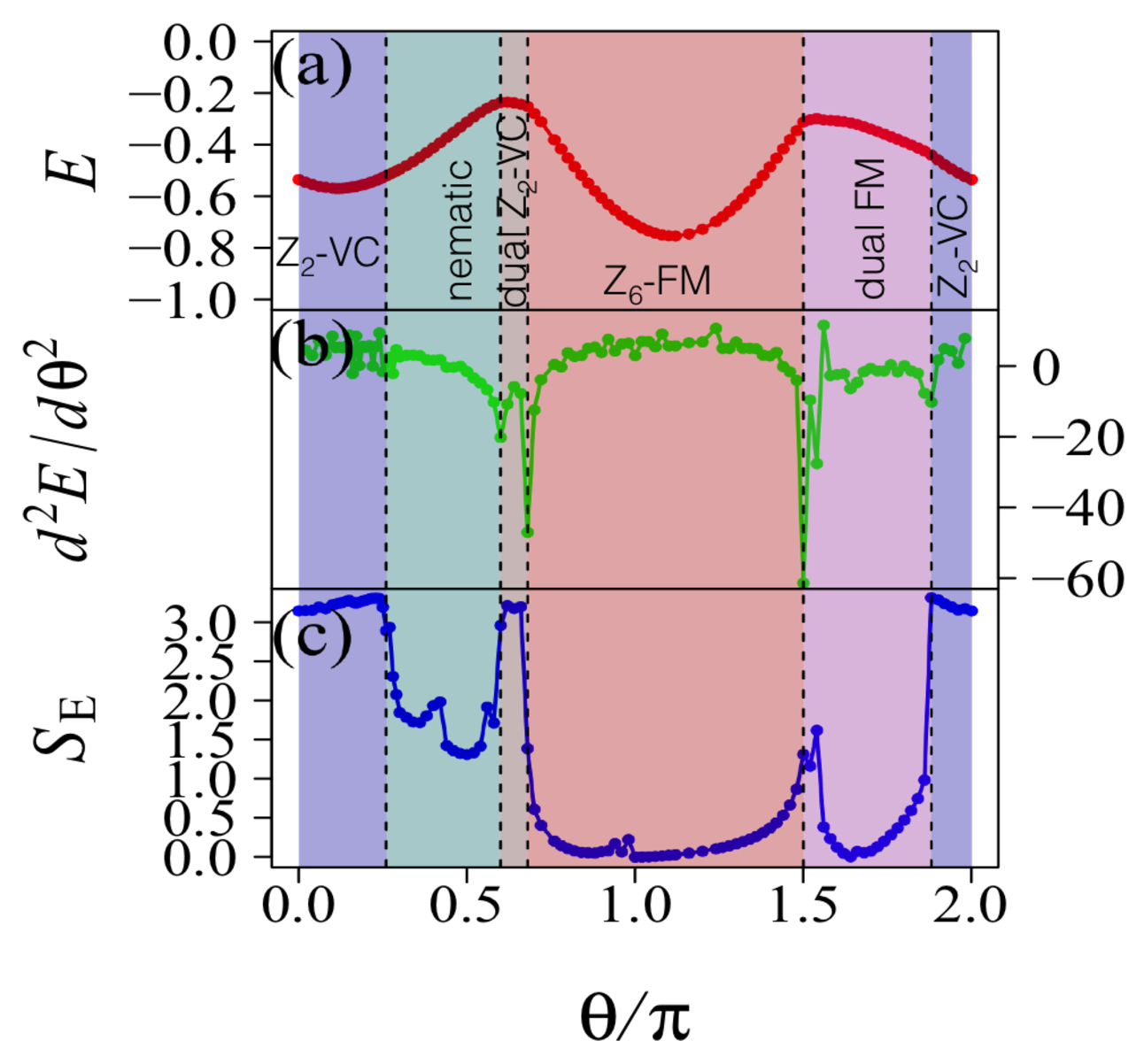}
          \hspace{1cm}
          \vspace{0cm}
        \end{center}
      \end{minipage}
      \\
      \begin{minipage}{1\hsize}
        \begin{center}
        \vspace{-0.5cm}
          \includegraphics[clip, width=20pc]{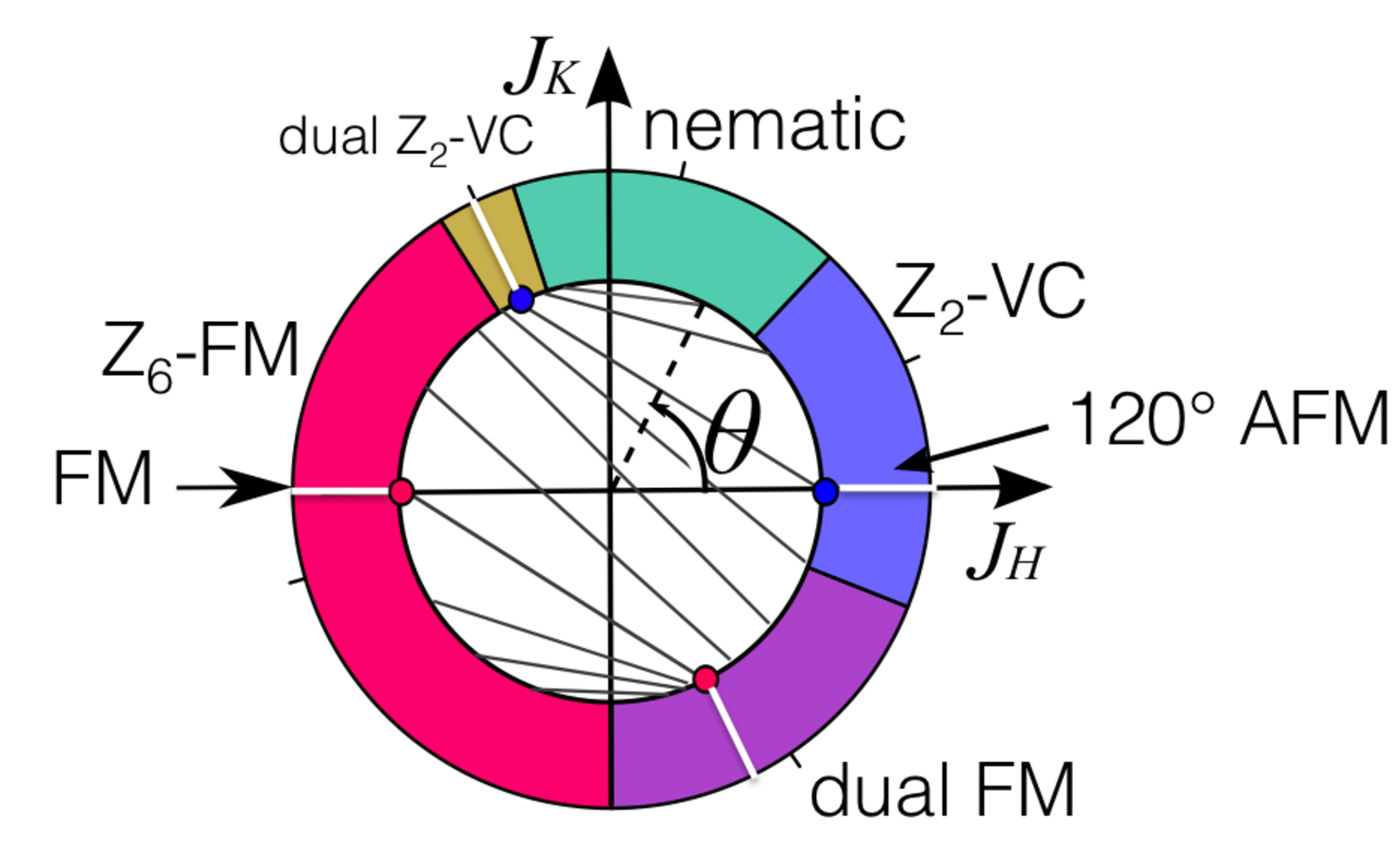}
          \hspace{0cm} (d)
          \vspace{0cm}
        \end{center}
      \end{minipage}

    \end{tabular}
    \caption{(Color Online) (a) Ground-state energy per site, $E$, of the KH model in Eq.~(\ref{KHmodel}) (red plots) obtained by the DMRG method for a 12$\times$6-site lattice with toroidal boundary conditions. There are a 120$^\circ$ AFM phase (only at $\theta=0$), a $\mathbb{Z}_2$ vortex crystal phase ($\mathbb{Z}_2$-VC), its dual phase (dual $\mathbb{Z}_2$-VC), a nematic phase (nematic), a $\mathbb{Z}_6$ ferromagnetic phase ($\mathbb{Z}_6$-FM), and its dual phase (dual FM). (b) Second derivative of $E$ with respect to $\theta$, $d^2E/d\theta ^2$ (green plots). (c) Entanglement entropy (blue plots) for a 12$\times$6-site lattice with the cylindrical boundary conditions. The vertical dotted lines denote the phase boundary determined by the second derivative of $E$. (d) Phase diagram of the KH model in Eq.~(\ref{KHmodel}). The circle parametrization of $J_H = \cos \theta$ and $J_K = \sin \theta$ is used. Gray lines inside the inner circle connect the parameter space related by the Klein duality. Filled blue and red circles on the inner circle show parameter points with SU(2) spin-rotational symmetry.}
    \label{phase}
  \end{center}
\end{figure}

For convenience, we take $J_K=\sin \theta$ and $J_H= \cos \theta$ in Eq.~(\ref{KHmodel}) using angle parameter $\theta$ [$\theta \in [0,2\pi)$].
Performing DMRG calculations for the 72-site lattice with the cylindrical boundary condition, we construct the phase diagram of the KH model as a function of $\theta$.
In Fig.~\ref{phase}(a), the ground-state energy per site, $E$, is plotted, and by analyzing the energy, a ground-state phase diagram is obtained as shown in Fig.~\ref{phase}(d).
Vertical dashed lines in Figs.~\ref{phase}(a)-\ref{phase}(c) indicate the positions of phase transitions determined by the peaks of the second derivative of $E$ with respect to $\theta$ shown in Fig.~\ref{phase}(b), and the change in the spin-spin correlation (not shown).
We conclude that all the phase transitions are of first order since the spin-spin correlation discontinuously changes at the boundaries in spite of the unclear discontinuity of the first derivative of energy.
For example, although the singular behavior of $d^2E/d\theta^2$ is unclear at $\theta = 0.26\pi$, the spin-spin correlation changes discontinuously, and thus we judge that a first-order phase transition occurs at $\theta =0.26\pi$.
The phase boundary of our DMRG result is close to the boundary of classical results but slightly different from ED results~\cite{rousochatzakis2012, becker2015}.
The difference between the ED and DMRG results, both of which include quantum effects, may be due to geometrical spin frustration and the existence of ordered phases with large unit cells, which make the size effect strong.

The phases denoted in Fig.~\ref{phase}(a) are identified by the spin-spin correlation functions and the spin structure factors (see Sect.~\ref{sec3.2}): they are a 120$^\circ$ AFM phase at $\theta=0$, a $\mathbb{Z}_2$-vortex crystal phase for $-0.15\pi<\theta<0.25\pi$, its dual counterpart, i.e., a dual $\mathbb{Z}_2$-vortex crystal phase for $0.62\pi<\theta<0.7\pi$, a nematic phase for $0.25\pi<\theta<0.62\pi$, a $\mathbb{Z}_6$ FM phase for $0.7\pi<\theta<1.5\pi$, and its counterpart, i.e., a dual FM phase, for $1.5\pi<\theta<1.85\pi$. The phases are fully consistent with previous studies~\cite{rousochatzakis2012, becker2015}, and no new phase is found, at least for systems up to 72 sites.

\subsection{Spin structure factors}
\label{sec3.2}
Next, we discuss magnetic phases in the phase diagram in detail.
We calculate the spin structure factors for each spin component given by
\begin{align}
\begin{split}
S^{\gamma \gamma} (\bvec{q})&=\sum _{l=0}^{N-1} \langle \hat S_0^\gamma \hat S_l ^\gamma \rangle \cos(\bvec{q} \cdot \bvec{r}_l), \\
S(\bvec{q})&=\sum _{\gamma =x,y,z} S^{\gamma \gamma}(\bvec{q}),
\end{split}
\end{align}
where $N$ is the total number of sites, $\bvec{r}_l$ denotes the position of the $l$th site, and $\bvec{q}=(q_x,q_y)$ is the wave number defined in the first Brillouin zone (BZ).
To calculate $S^{\gamma \gamma} (\bvec{q})$, we use the toroidal boundary condition for the 72-site lattice.

\subsubsection{120$^\circ$ AFM and $\mathbb{Z}_2$-vortex crystal phases}
\label{sec3.2.1}

\begin{figure}[t]
  \begin{center}
    \begin{tabular}{r}
      \begin{minipage}{0.5\hsize}
        \begin{center}
          \includegraphics[clip, width=10pc]{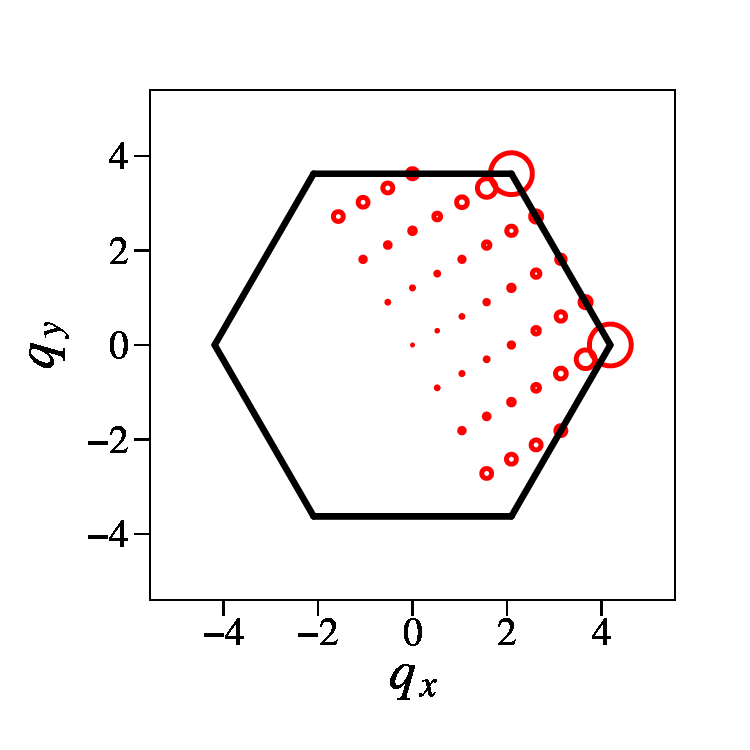}
          \hspace{1cm} (a)
          \vspace{0cm}
        \end{center}
      \end{minipage}
      \begin{minipage}{0.5\hsize}
        \begin{center}
          \includegraphics[clip, width=10pc]{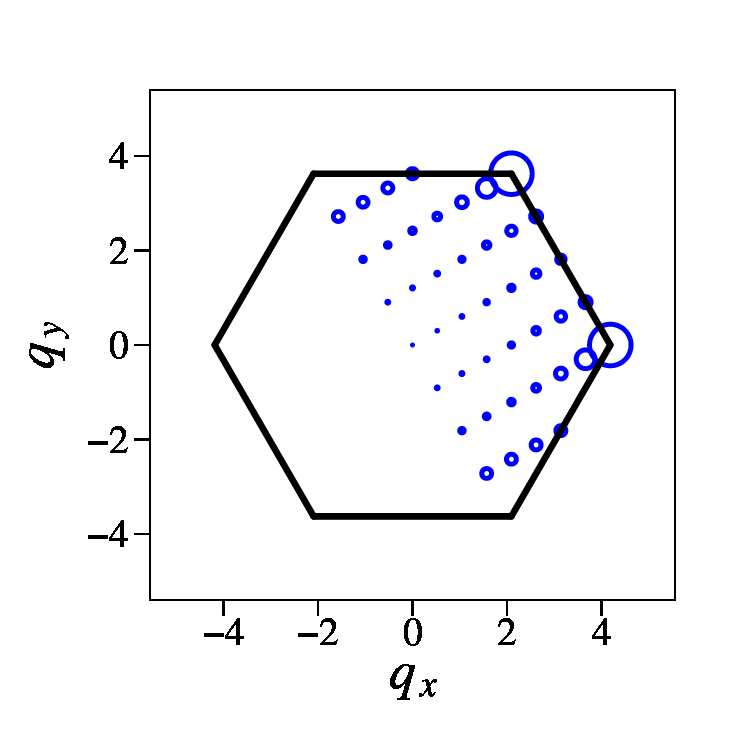}
          \hspace{1cm} (b)
          \vspace{0cm}
        \end{center}
      \end{minipage}
      \\
      \begin{minipage}{0.5\hsize}
        \begin{center}
          \includegraphics[clip, width=10pc]{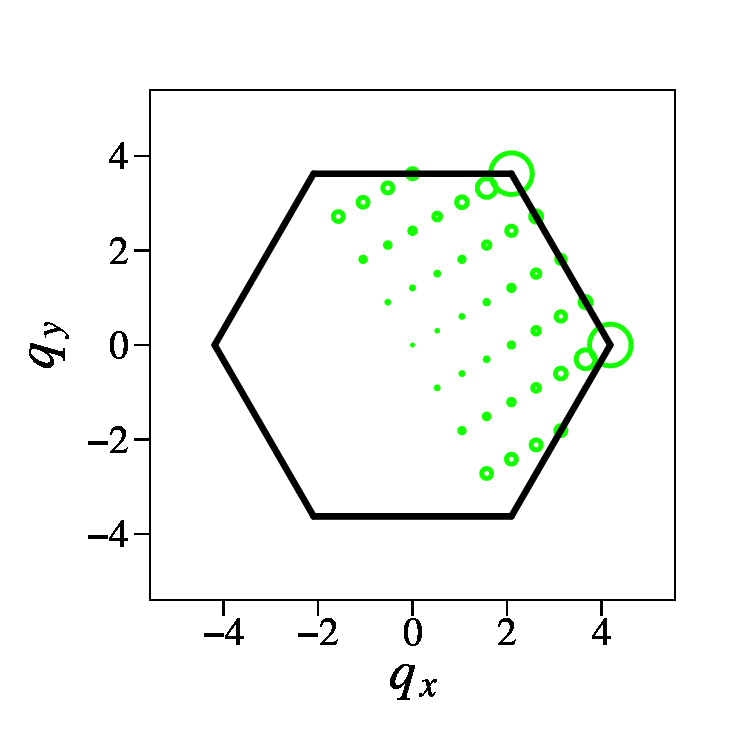}
          \hspace{1cm} (c)
          \vspace{0cm}
        \end{center}
      \end{minipage}
      \begin{minipage}{0.5\hsize}
        \begin{center}
          \includegraphics[clip, width=10pc]{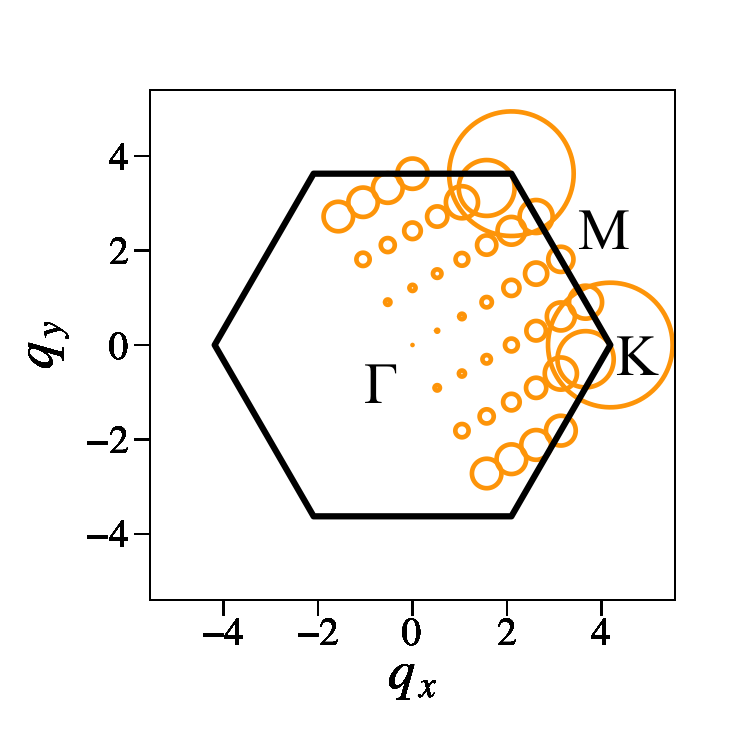}
          \hspace{1cm} (d)
          \vspace{0cm}
        \end{center}
      \end{minipage}
    \end{tabular}
    \caption{(Color Online) Spin structure factors at $\theta =0$ (120$^\circ$ AFM phase) for a 12$\times$6-site lattice with the toroidal boundary conditions. (a) $S^{xx}(\bvec{q})$, (b) $S^{yy}(\bvec{q})$, (c) $S^{zz}(\bvec{q})$, and (d) $S(\bvec{q})$. The radius of the circles represents the intensity of the structure factors at the momentum point defined by the lattice. The results in half of the first BZ are plotted. }
    \label{sq_0.00}
  \end{center}
\end{figure}

First, we examine the magnetic structures of the 120$^\circ$ AFM ordered phase ($\theta =0$) and the $\mathbb{Z}_2$-vortex crystal phase ($-0.15\pi<\theta<0.25\pi$).
Figure~\ref{sq_0.00} shows the spin structure factors at $\theta =0$, where the results in half of the first BZ are plotted and the radius of colored circles represents the intensity of the structure factors.
Because of the SU(2) symmetry, the three components, $S^{xx}(\bvec{q})$, $S^{yy}(\bvec{q})$, and $S^{zz}(\bvec{q})$, are equivalent to each other.
The largest intensity appears at the $K$ points in the first BZ, corresponding to a 120$^\circ$ AFM order.

\begin{figure}[t]
  \begin{center}
    \begin{tabular}{r}
      \begin{minipage}{0.5\hsize}
        \begin{center}
          \includegraphics[clip, width=10pc]{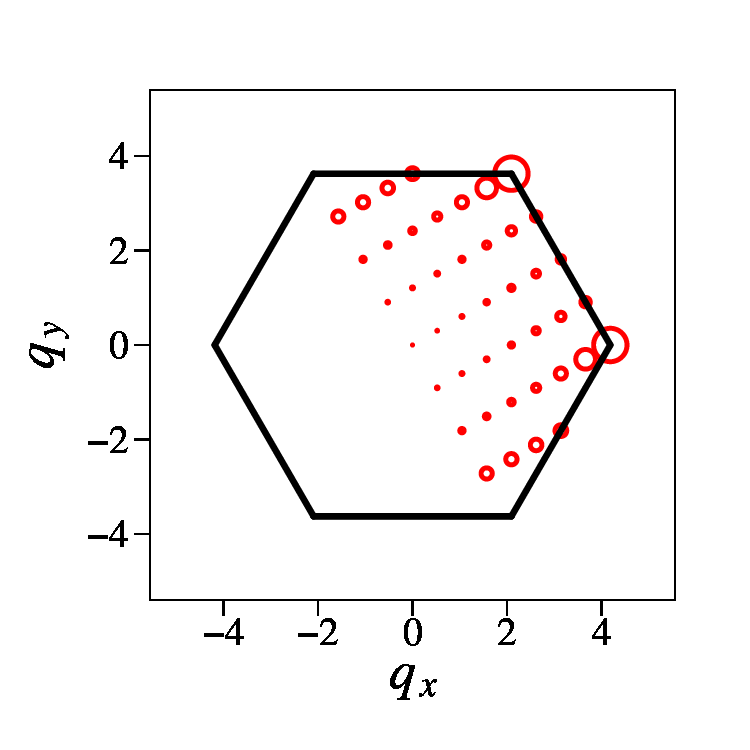}
          \hspace{1cm} (a)
          \vspace{0cm}
        \end{center}
      \end{minipage}
      \begin{minipage}{0.5\hsize}
        \begin{center}
          \includegraphics[clip, width=10pc]{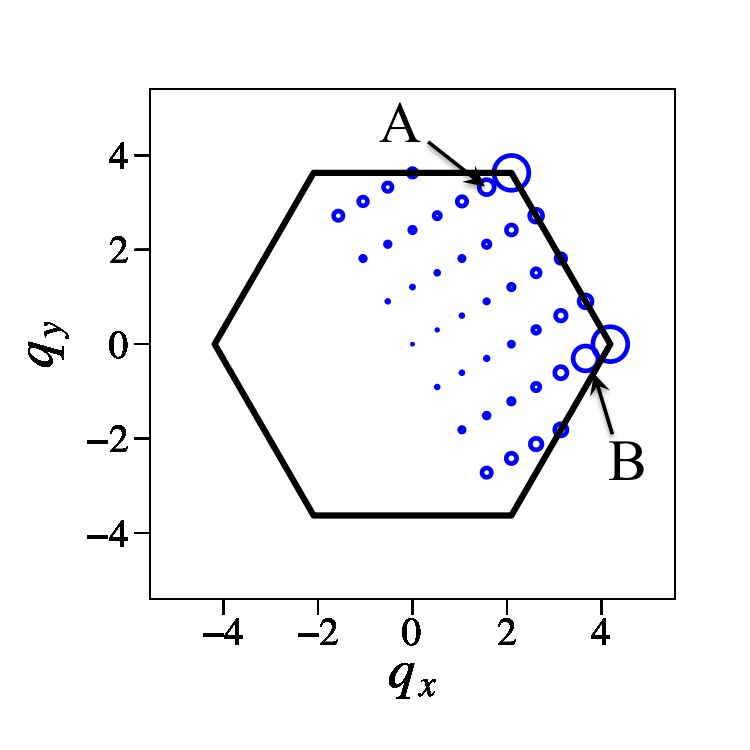}
          \hspace{1cm} (b)
          \vspace{0cm}
        \end{center}
      \end{minipage}
      \\
      \begin{minipage}{0.5\hsize}
        \begin{center}
          \includegraphics[clip, width=10pc]{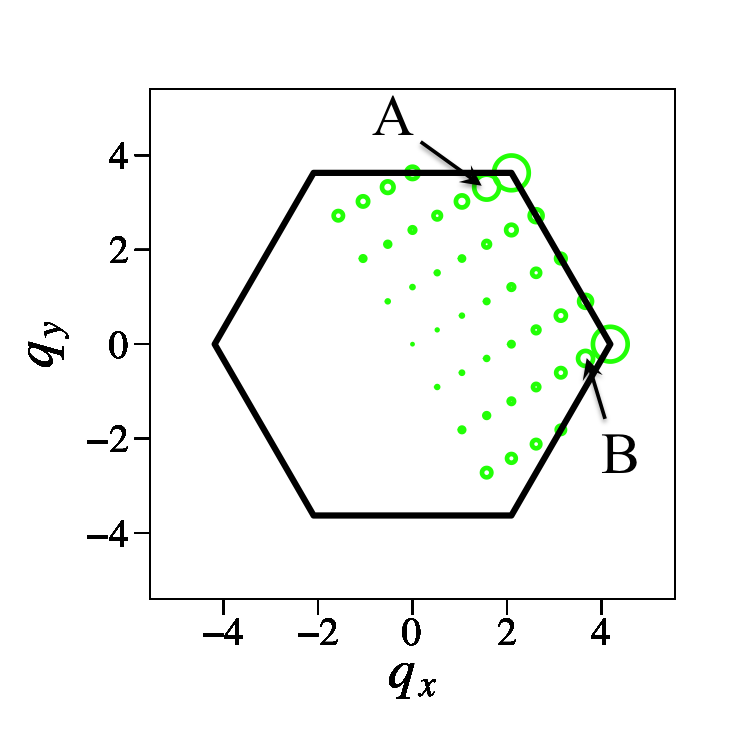}
          \hspace{1cm} (c)
          \vspace{0cm}
        \end{center}
      \end{minipage}
      \begin{minipage}{0.5\hsize}
        \begin{center}
          \includegraphics[clip, width=10pc]{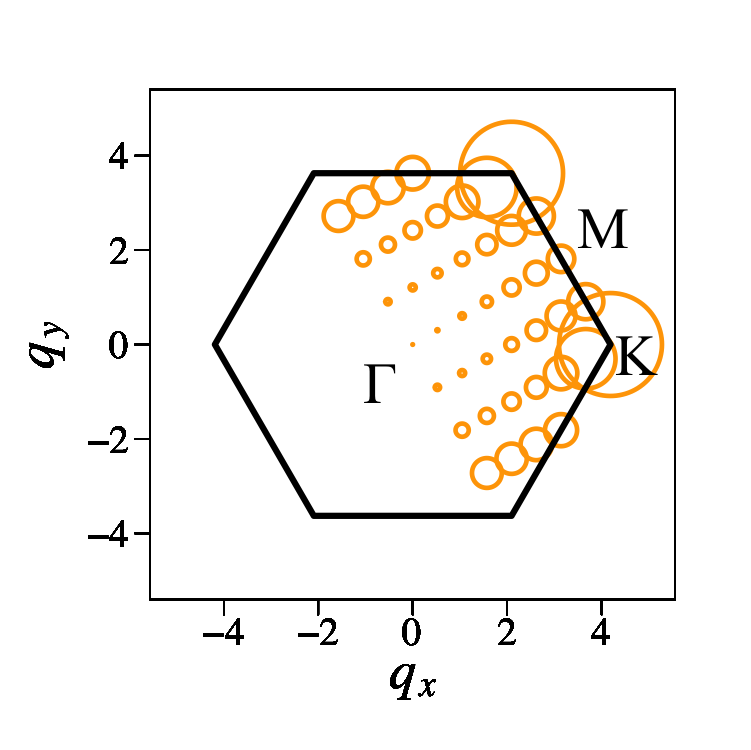}
          \hspace{1cm} (d)
          \vspace{0cm}
        \end{center}
      \end{minipage}
    \end{tabular}
    \caption{(Color Online) Same as Fig.~\ref{sq_0.00} but for $\theta =0.16\pi$ in $\mathbb{Z}_2$-vortex crystal phase. The symbols A and B in (b) and (c) denote the momenta where asymmetric behavior with respect to the $\Gamma$-M line is clearly seen.}
    \label{sq_0.16}
  \end{center}
\end{figure}

When Kitaev couplings are introduced, the SU(2) symmetry is broken, which will lead to anisotropic behavior in the spin structure factors.
Such behavior is actually shown in Fig.~\ref{sq_0.16}, where $\theta = 0.16\pi$.
It is clear that $S^{yy}(\bvec{q})$ and $S^{zz}(\bvec{q})$ in Figs.~\ref{sq_0.16} (b) and \ref{sq_0.16} (c), respectively, show different values at the two points A and B (see the figures), whereas the structure factors are exactly the same at $\theta=0$.
On the other hand, $S^{xx}(\bvec{q})$ exhibits the same intensity at the A and B points.
This implies that two of the three spin components are asymmetric with respect to a symmetric line connecting the $\Gamma$ point [$\bvec{q}=(0,0)$] and the midpoint of two equivalent K points, i.e., the M point.
In our calculation, $S^{xx}(\bvec{q})$ is selected as a symmetric component with respect to the $\Gamma$-M line because of the assignment of the $xx$-bond direction parallel to the line (see Fig.~\ref{triangle}).
To be more precise, we find that $S^{yy}(\bvec{q})$ is smaller at A than at B, and vice versa for $S^{zz}(\bvec{q})$, i.e., $(S^{yy}, S^{zz})=(0.47,0.79)$ at A and $(S^{yy}, S^{zz})=(0.79, 0.47)$ at B. This tendency is also true, for example, at $\theta =0.08\pi$, where $(S^{yy}, S^{zz})=(0.53, 0.67)$ at A and $(S^{yy}, S^{zz})=(0.67, 0.53)$ at B.

The asymmetry with respect to the $\Gamma$-M line is consistent with the spin structure factor proposed for the $\mathbb{Z}_2$-vortex crystal phase~\cite{rousochatzakis2012, becker2015}, where an ordering wave vector shifts from the K points with accompanying secondary Fourier components away from K.
As a result, a finite Kitaev coupling breaks the 6-fold rotational symmetry of $S^{xx}(\bvec{q})$, $S^{yy}(\bvec{q})$, and $S^{zz}(\bvec{q})$ around the $\Gamma$ point down to 2-fold rotational symmetry, although the 6-fold rotational symmetry remains in $S(\bvec{q})=S^{xx}(\bvec{q})+S^{yy}(\bvec{q})+S^{zz}(\bvec{q})$.

\subsubsection{Nematic phase}
\label{sec3.2.2}

It has been suggested that a nematic phase appears around $\theta =0.5\pi$~\cite{rousochatzakis2012,becker2015}.
The magnetic order in the nematic phase is known to be of the AFM Ising-chain-type with one of the three spin components ($\gamma$) ordered in the corresponding lattice direction $\gamma\gamma$~\cite{rousochatzakis2012}.
Since the other directions are disconnected in terms of the given spin component, flipping all the spins of the Ising chain does not change its energy.
This leads to sub-extensive degeneracy of the ground state, which is related to an intermediate symmetry lying midway between the global symmetries and local gauge symmetries~\cite{becker2015, jackeli2015}.
The intermediate symmetry and dimensional reduction have been discussed on the basis of the compass model, indicating that a higher-dimensional spin system decouples into lower-dimensional subsystems.
The low-energy states are ordered along only one direction in the system with intermediate symmetry, leading to degeneracy.
Such degeneracy can be lifted by thermal or quantum fluctuations by the so-called order-by-disorder mechanism~\cite{jackeli2015}.
Such a partial order composed of decoupled chains also appears in the compass model of a square lattice.

\begin{figure}[t]
  \begin{center}
    \begin{tabular}{r}
      \begin{minipage}{0.5\hsize}
        \begin{center}
          \includegraphics[clip, width=10pc]{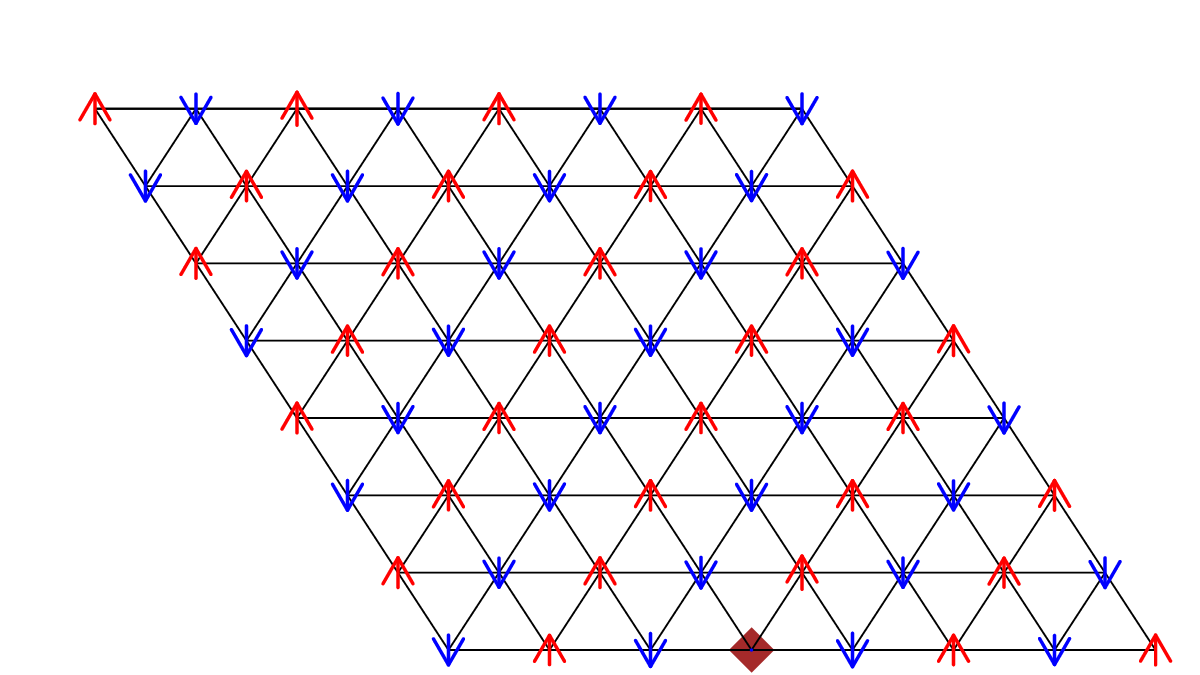}
          \hspace{1cm} (a)
          \vspace{0cm}
        \end{center}
      \end{minipage}
      \begin{minipage}{0.5\hsize}
        \begin{center}
          \includegraphics[clip, width=10pc]{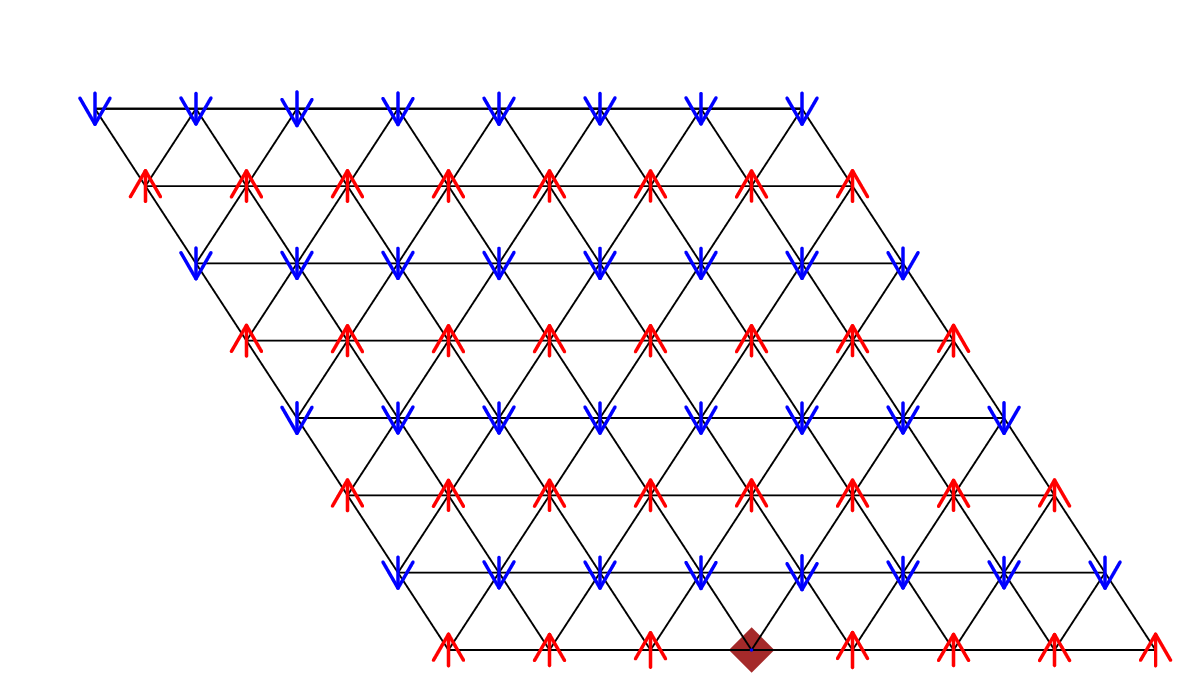}
          \hspace{1cm} (b)
          \vspace{0cm}
        \end{center}
      \end{minipage}
    \end{tabular}
    \caption{(Color Online) $\langle S_i^x S_j^x \rangle$ for the nematic phase at $\theta =0.5\pi$ in an $8\times8$ lattice with the toroidal boundary conditions. The energies of (a) and (b) are degenerate. Site $i$ is indicated by a brown diamond. Upward red arrows and downward blue arrows show positive and negative values of the spin-spin correlation, respectively. The length of the arrows represents the strength of the spin-spin correlation. The $x$ component of the spins aligns antiferromagnetically and ferromagnetically along the horizontal direction of the lattice in (a) and (b), respectively.}
    \label{spinspin-8x8p}
  \end{center}
\end{figure}

In the nematic phase at $\theta =0.5\pi$, i.e., the AFM Kitaev point, it has been pointed out that the AFM correlation between next-nearest-neighbor chains, which locks the next-nearest-neighbor spin alignment, reduces the sub-extensive degeneracy of the ground state from $3\times2^L$ to the nonextensive value $3\times2^2$, where $L$ denotes the total number of chains~\cite{becker2015}.
This result was obtained from DMRG calculations with three-leg and four-leg triangular ladders for $L$ up to $14$~\cite{becker2015}.
We have confirmed this reduction of the degeneracy to $3\times2^2$ for a $6\times4$ lattice.
In order to obtain $2^2$-fold degeneracy for one component, we apply a small external magnetic field at two sites on neighboring chains to force the direction of neighboring spins to be parallel or antiparallel, and find that the energies of the two cases are the same within the numerical accuracy.
In contrast to the degeneracy seen in the $6\times4$ lattice, we find that a larger system of $12\times6$ does not show such degeneracy: the energy with an antiparallel spin alignment for the neighboring chains along the $xx$-bond direction is always lower than that with a parallel spin alignment forced by a small magnetic field.
This is probably due to the anisotropic geometry of the $12\times6$ lattice as compared with the $6\times4$ lattice, resulting in the AFM alignment of a given component of spins along the horizontal direction (the $xx$-bond direction) of the $12\times6$ lattice.

In order to reduce this anisotropic geometry, we additionally examine an $8\times8$ square lattice under the toroidal boundary conditions.
As expected, we find degenerate ground states with the same energy by applying a small magnetic field.
Figure~\ref{spinspin-8x8p} shows the spin-spin correlation function for the $x$ component between sites $i$ and $j$ of the $8\times8$ lattice, given by $\langle S_i^x S_j^x \rangle = \langle 0| \hat S_i^x\hat S_j^x |0\rangle$, where $|0\rangle$ is the ground-state wave function and site $i$ is chosen to be a site denoted by the diamond in the figure.
In Fig.~\ref{spinspin-8x8p}(a), we notice that the $x$ component forms an AFM chain along the $xx$-bond direction (see Fig.~\ref{triangle}) and a pair of sites connecting next-nearest-neighbor chains with the shortest distance exhibits the AFM alignment of the $x$ component.
These features are the same as in Fig.~\ref{spinspin-8x8p}(b) but the alignment between nearest-neighbor chains is different.
In other words, the $x$ component of spins aligns antiferromagnetically along the horizontal direction of the $8\times8$ lattice in Fig.~\ref{spinspin-8x8p}(a) but aligns ferromagnetically in Fig.~\ref{spinspin-8x8p}(b).
This is consistent with the presence of $3\times2^2$-fold degeneracy.

\begin{figure}[t]
  \begin{center}
    \begin{tabular}{r}
      \begin{minipage}{0.5\hsize}
        \begin{center}
          \includegraphics[clip, width=10pc]{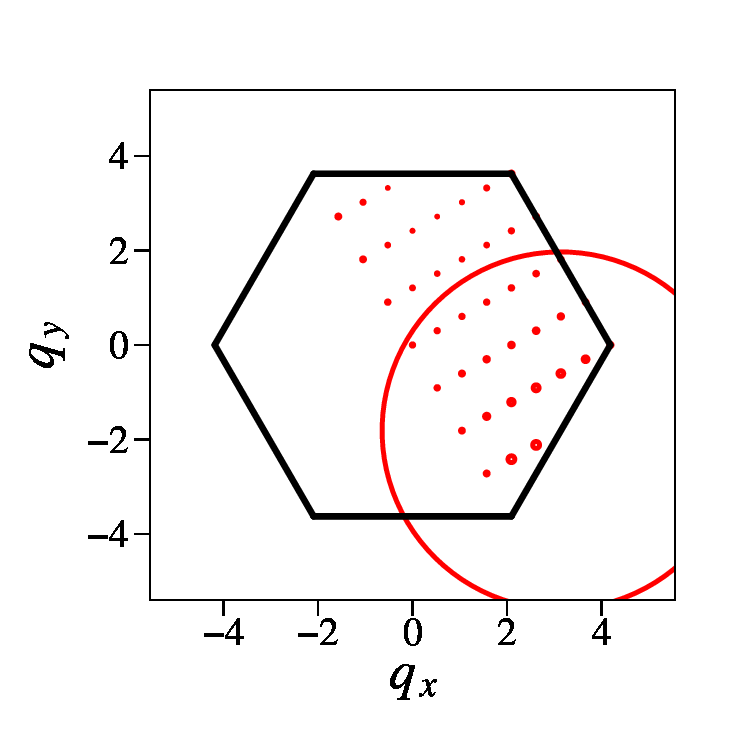}
          \hspace{1cm} (a)
          \vspace{0cm}
        \end{center}
      \end{minipage}
      \begin{minipage}{0.5\hsize}
        \begin{center}
          \includegraphics[clip, width=10pc]{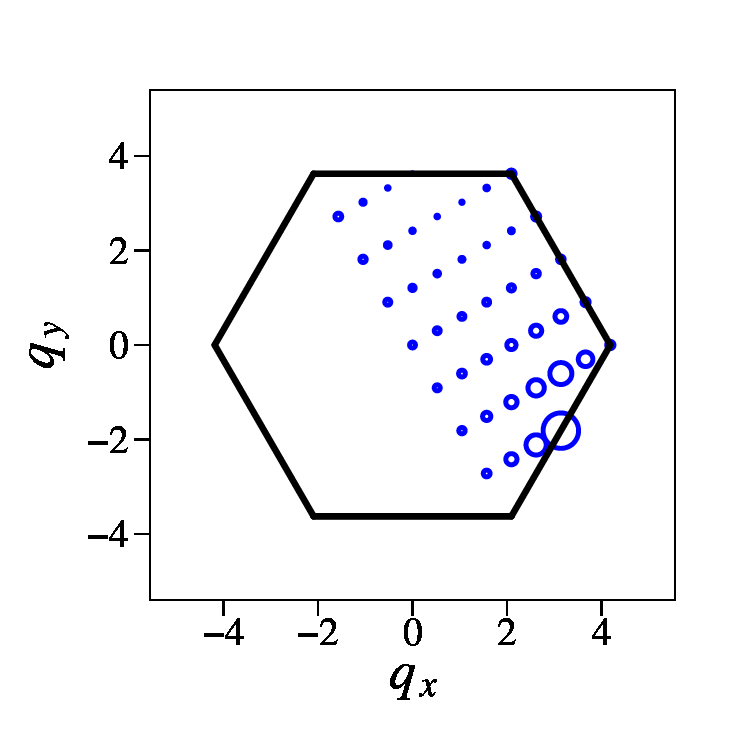}
          \hspace{1cm} (b)
          \vspace{0cm}
        \end{center}
      \end{minipage}
      \\
      \begin{minipage}{0.5\hsize}
        \begin{center}
          \includegraphics[clip, width=10pc]{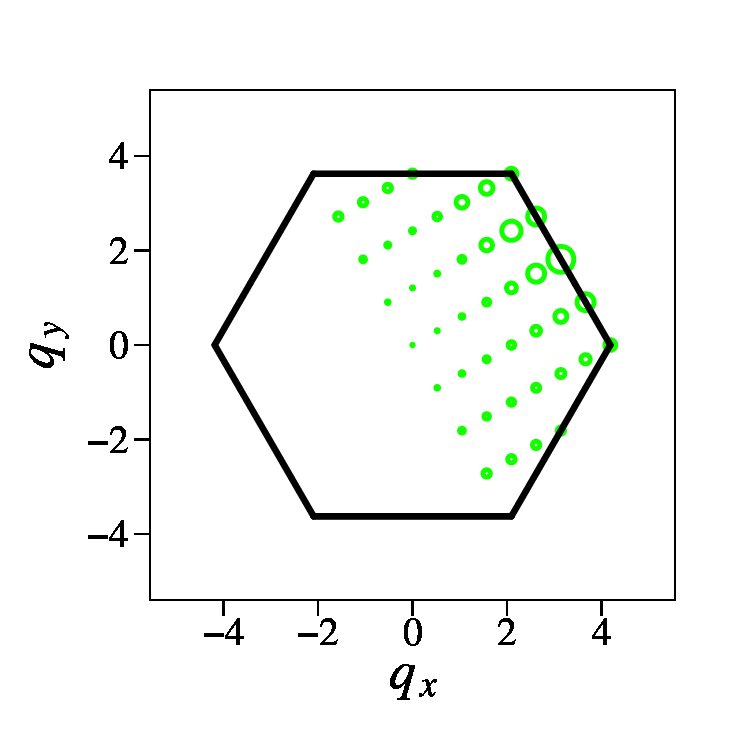}
          \hspace{1cm} (c)
          \vspace{0cm}
        \end{center}
      \end{minipage}
      \begin{minipage}{0.5\hsize}
        \begin{center}
          \includegraphics[clip, width=10pc]{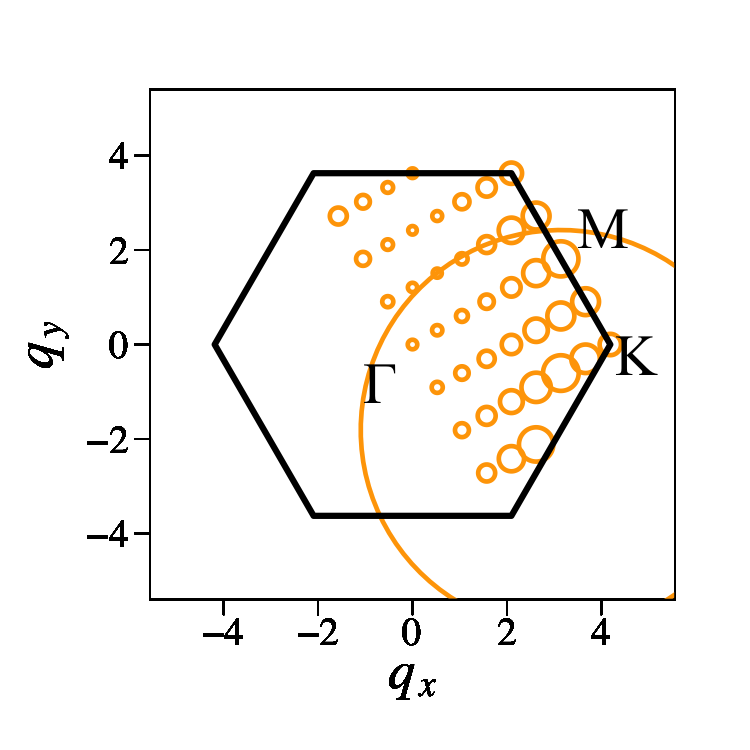}
          \hspace{1cm} (d)
          \vspace{0cm}
        \end{center}
      \end{minipage}
    \end{tabular}
    \caption{(Color Online) Same as Fig.~\ref{sq_0.00} but for $\theta =0.48\pi$ in nematic phase. The center of the large circle in (a) and (d) is located at the M point with negative $q_y$.}
    \label{sq_0.48}
  \end{center}
\end{figure}

Away from the AFM Kitaev point, it has been pointed out that the degeneracy is further reduced to $3\times2$~\cite{becker2015} even in the nematic phase.
When an AFM Heisenberg interaction is added within the nematic phase, i.e., $\theta<0.5\pi$, it is expected that the ground state with the AFM spin alignment between nearest-neighbor chains such as that shown in Fig.~\ref{spinspin-8x8p}(a) will be selected.
The $12\times6$ lattice actually has such a ground state.
The spin structure factors for $\theta=0.48\pi$ are shown in Fig.~\ref{sq_0.48}, where the $x$ component is selected in contrast to other components.
$S^{xx}(\bvec{q})$ has the largest value at the M point with negative $q_y$ as expected from the spin structure shown in Fig.~\ref{spinspin-8x8p}(a).

\subsubsection{$\mathbb{Z}_6$ ferromagnetic phase}
\label{sec3.2.3}

\begin{figure}[t]
  \begin{center}
    \begin{tabular}{r}
      \begin{minipage}{0.5\hsize}
        \begin{center}
          \includegraphics[clip, width=10pc]{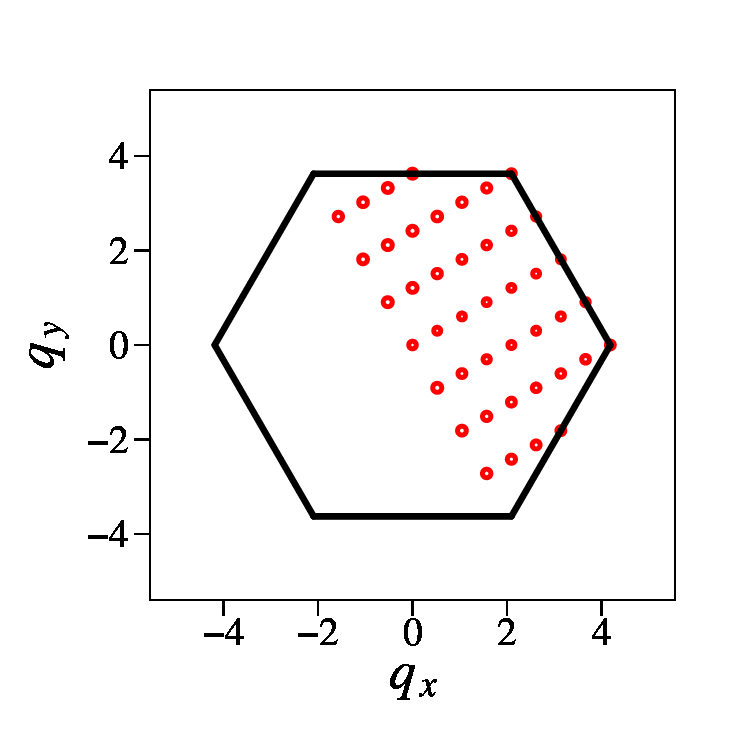}
          \hspace{1cm} (a)
          \vspace{0cm}
        \end{center}
      \end{minipage}
      \begin{minipage}{0.5\hsize}
        \begin{center}
          \includegraphics[clip, width=10pc]{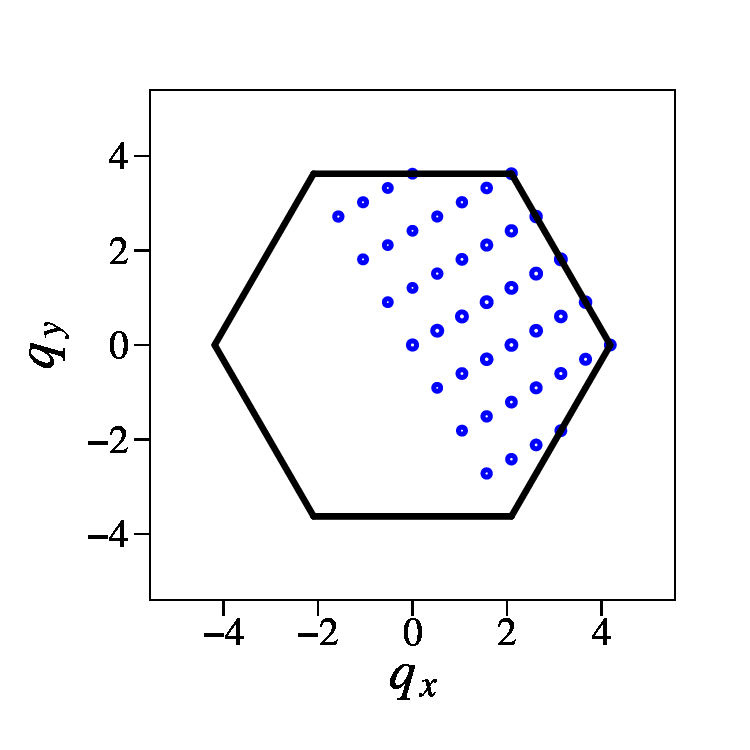}
          \hspace{1cm} (b)
          \vspace{0cm}
        \end{center}
      \end{minipage}
      \\
      \begin{minipage}{0.5\hsize}
        \begin{center}
          \includegraphics[clip, width=10pc]{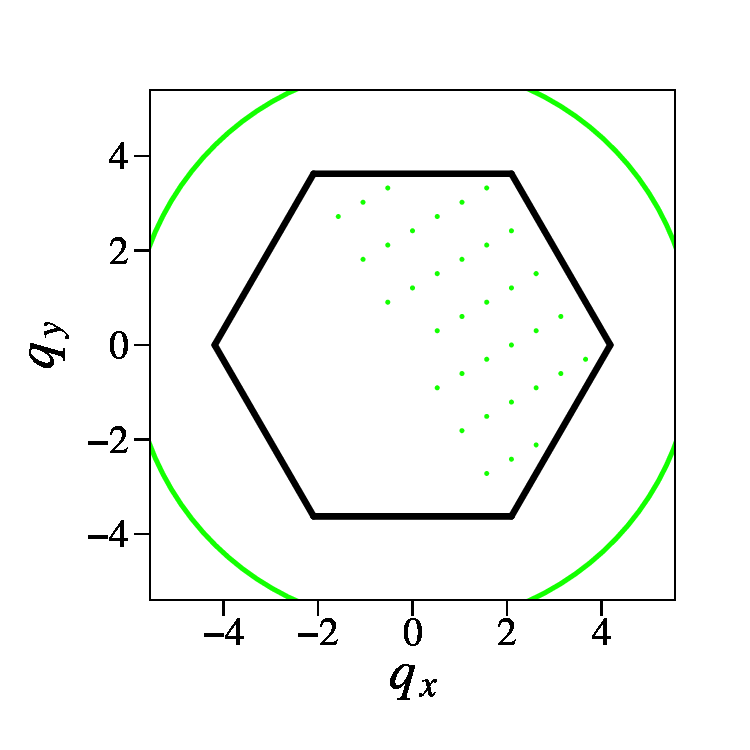}
          \hspace{1cm} (c)
          \vspace{0cm}
        \end{center}
      \end{minipage}
      \begin{minipage}{0.5\hsize}
        \begin{center}
          \includegraphics[clip, width=10pc]{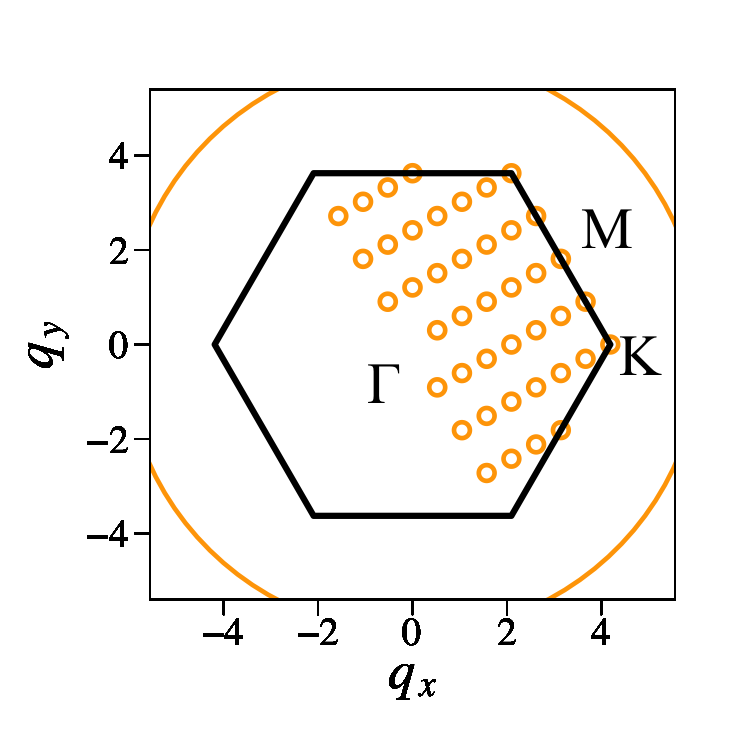}
          \hspace{1cm} (d)
          \vspace{0cm}
        \end{center}
      \end{minipage}
    \end{tabular}
    \caption{(Color Online) Same as Fig.~\ref{sq_0.00} but for $\theta =0.88\pi$ in $\mathbb{Z}_6$ FM phase. The center of the large circle in (c) and (d) is located at the $\Gamma$ point. }
    \label{sq_0.88}
  \end{center}
\end{figure}

\begin{figure}[htb]
  \begin{center}
    \begin{tabular}{r}
      \begin{minipage}{0.5\hsize}
        \begin{center}
          \includegraphics[clip, width=10pc]{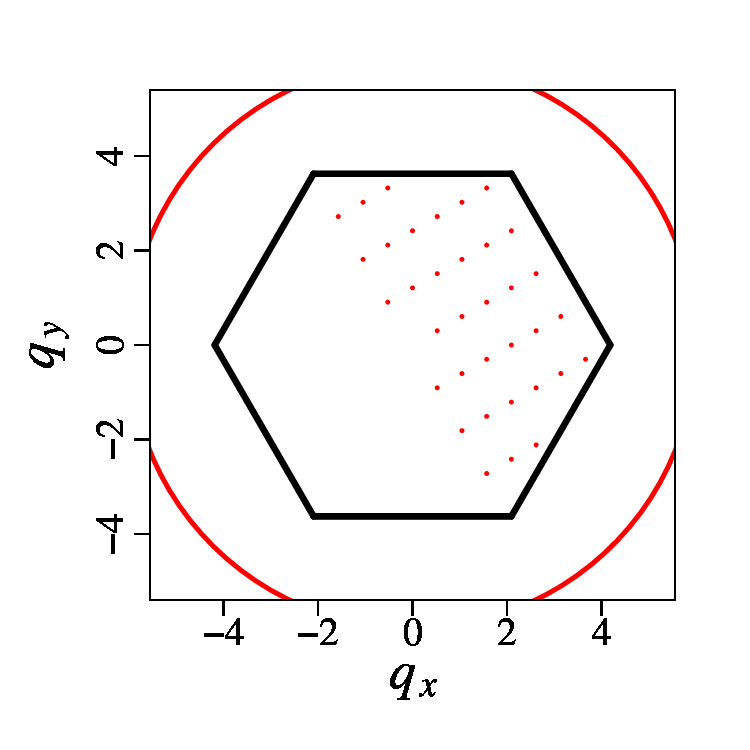}
          \hspace{1cm} (a)
          \vspace{0cm}
        \end{center}
      \end{minipage}
      \begin{minipage}{0.5\hsize}
        \begin{center}
          \includegraphics[clip, width=10pc]{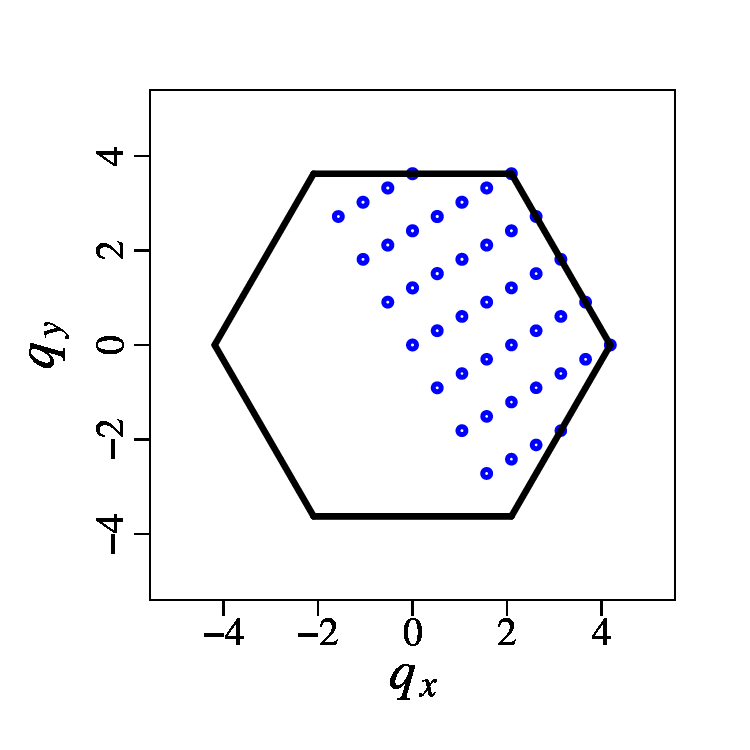}
          \hspace{1cm} (b)
          \vspace{0cm}
        \end{center}
      \end{minipage}
      \\
      \begin{minipage}{0.5\hsize}
        \begin{center}
          \includegraphics[clip, width=10pc]{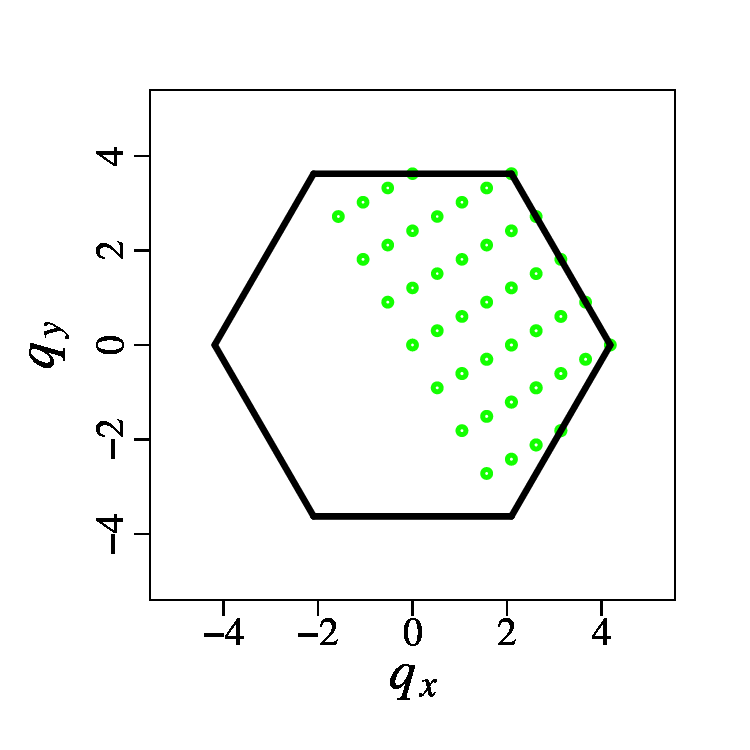}
          \hspace{1cm} (c)
          \vspace{0cm}
        \end{center}
      \end{minipage}
      \begin{minipage}{0.5\hsize}
        \begin{center}
          \includegraphics[clip, width=10pc]{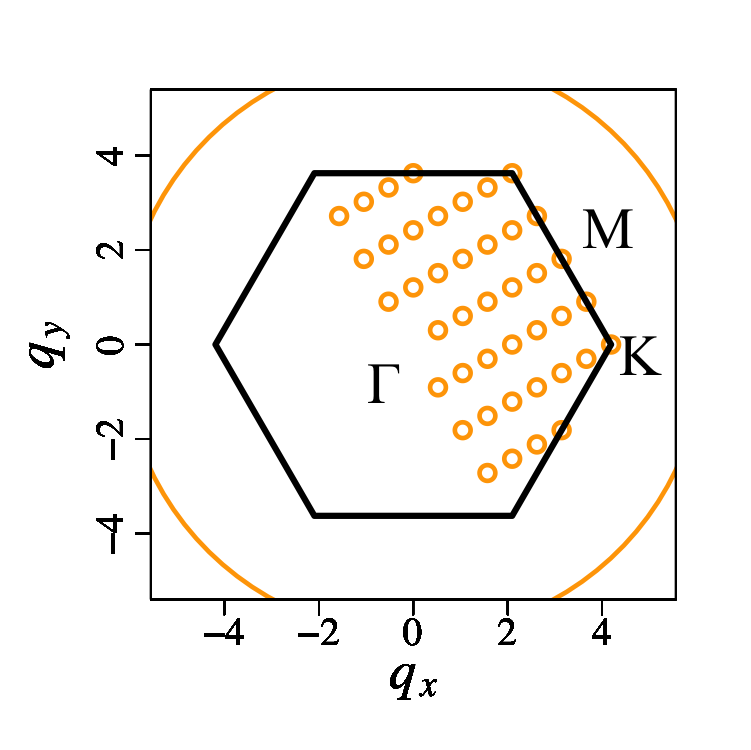}
          \hspace{1cm} (d)
          \vspace{0cm}
        \end{center}
      \end{minipage}
    \end{tabular}
    \caption{(Color Online) Same as Fig.~\ref{sq_0.00} but for $\theta =1.04\pi$ in $\mathbb{Z}_6$ FM phase. The center of the large circle in (a) and (d) is located at the $\Gamma$ point. }
    \label{sq_1.04}
  \end{center}
\end{figure}

The FM phase extends from $\theta =0.68\pi$ to $1.50\pi$.
A Heisenberg FM point is at $\theta =\pi$, where the Hamiltonian is SU(2) symmetric and the spin structure factors are the same for the $x$, $y$, and $z$ components.
The FM phase with finite Kitaev coupling is called the $\mathbb{Z}_6$ FM~\cite{becker2015} phase.
Although the order parameter for SU(2) symmetric systems takes its value on the whole unit sphere, Kitaev coupling discretizes it to $\mathbb{Z}_6$.
In our DMRG calculations, the $z$ ($x$) component of the spin configurations is selected when $0.68\pi<\theta<\pi$ ($\pi<\theta<1.50\pi$). For example, at $\theta=0.88\pi$ ($\theta=1.04\pi$) $S^{zz}(0,0)$ [$S^{xx}(0,0)$] has the largest intensity as shown in Fig.~\ref{sq_0.88} (Fig.~\ref{sq_1.04}).
The selection of a particular component of the spin structure factor is an artifact of the shape of the cluster and the choice of the one-dimensional path in the sweeping process of the DMRG method.
The three components should be degenerate.

\subsubsection{Duality}
\label{sec3.2.4}

It is well known that there is Klein duality on the parameter space of the KH model, which is given by $J_H \rightarrow -J_H$ and $J_K \rightarrow 2J_H + J_K$ as shown in Fig.~\ref{phase}(d)~\cite{kimchi2014}.
The duality is accompanied by spin rotations that depends on the sublattices~\cite{khaliullin2005}. 
Accordingly, two explicit SU(2) symmetric points at $\theta =0$ (AFM Heisenberg limit) and $\theta = \pi$ (FM Heisenberg limit) are mapped to two hidden SU(2) symmetric points at $\theta_0=\arctan(-2)=0.64758\pi$ and $\theta_\pi=\pi+\arctan(-2)=1.64758\pi$, respectively.

\begin{figure}[t]
  \begin{center}
    \begin{tabular}{r}
      \begin{minipage}{0.5\hsize}
        \begin{center}
          \includegraphics[clip, width=10pc]{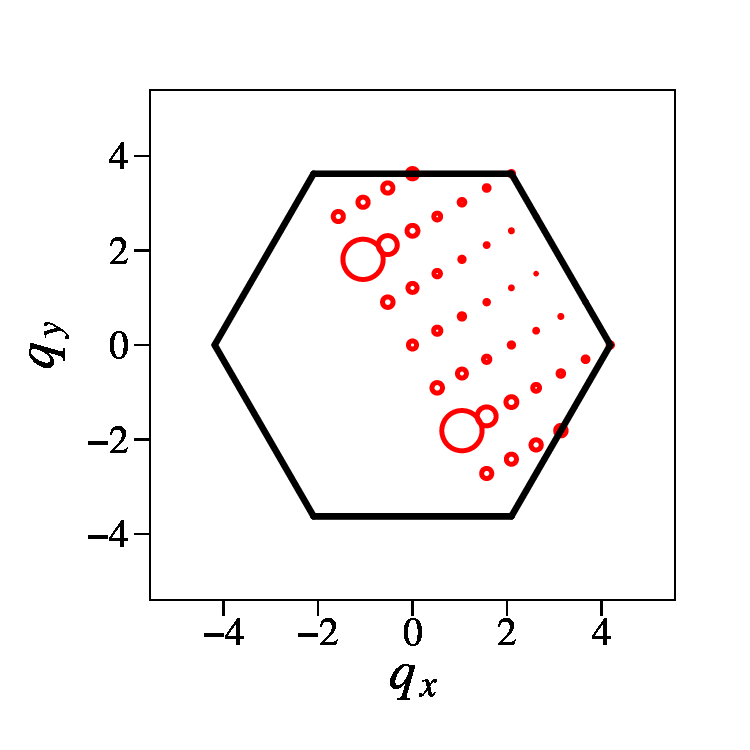}
          \hspace{1cm} (a)
          \vspace{0cm}
        \end{center}
      \end{minipage}
      \begin{minipage}{0.5\hsize}
        \begin{center}
          \includegraphics[clip, width=10pc]{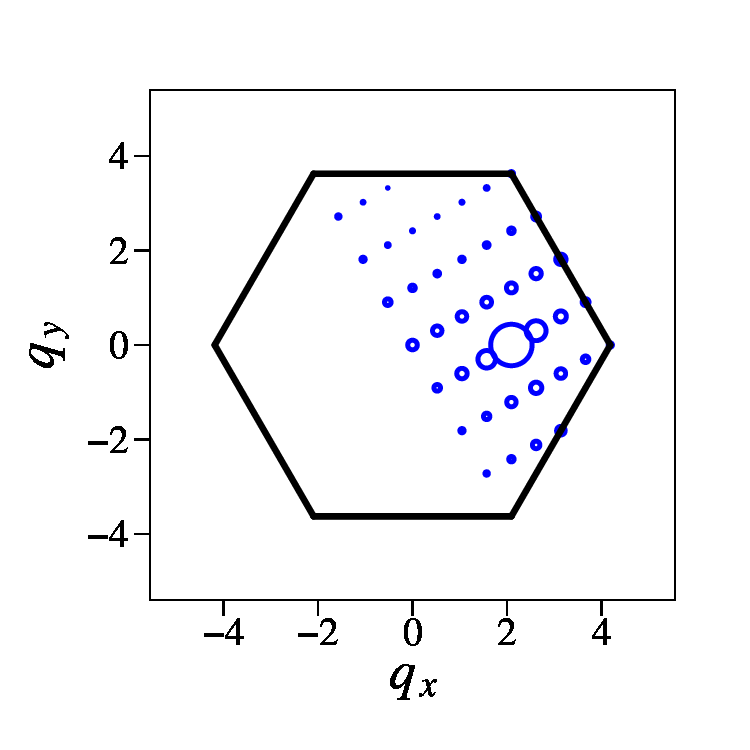}
          \hspace{1cm} (b)
          \vspace{0cm}
        \end{center}
      \end{minipage}
      \\
      \begin{minipage}{0.5\hsize}
        \begin{center}
          \includegraphics[clip, width=10pc]{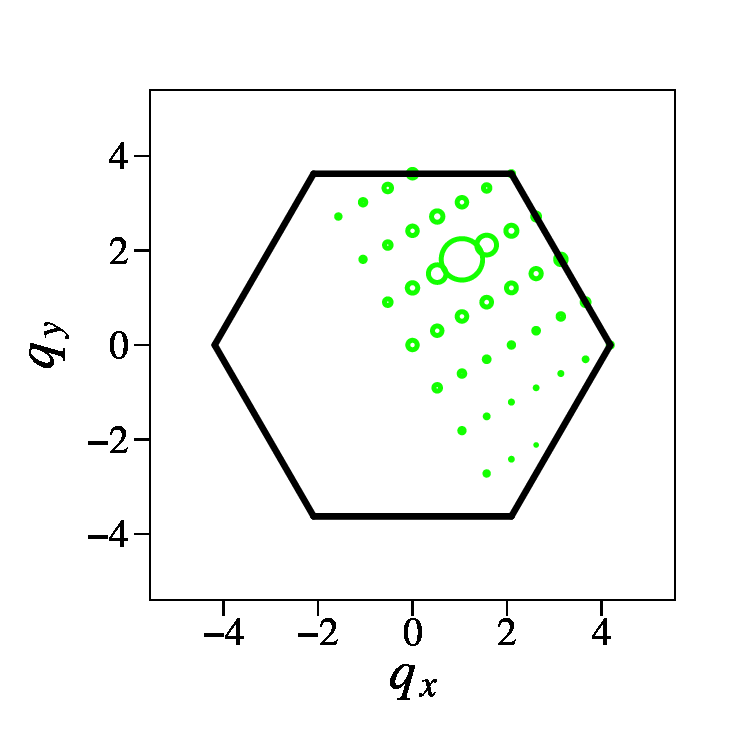}
          \hspace{1cm} (c)
          \vspace{0cm}
        \end{center}
      \end{minipage}
      \begin{minipage}{0.5\hsize}
        \begin{center}
          \includegraphics[clip, width=10pc]{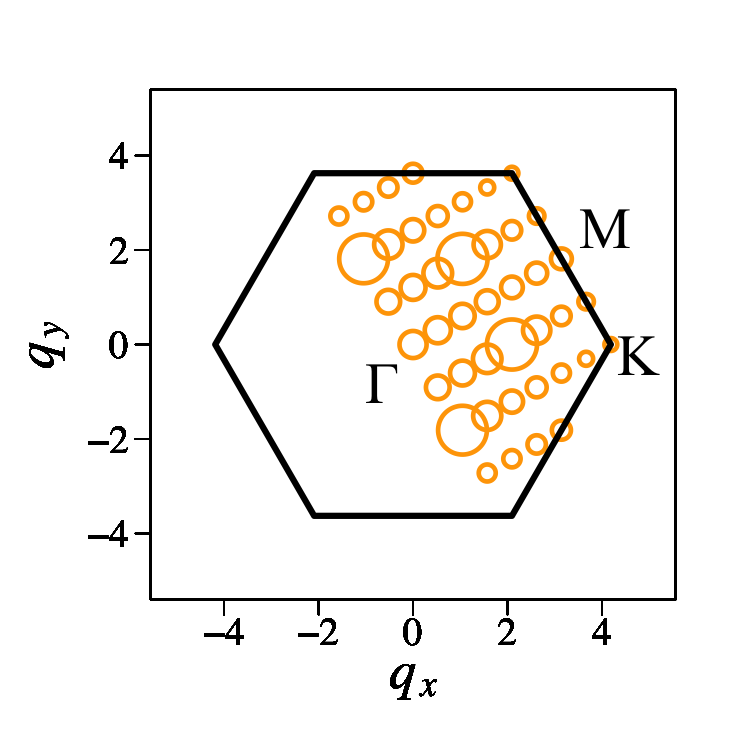}
          \hspace{1cm} (d)
          \vspace{0cm}
        \end{center}
      \end{minipage}
    \end{tabular}
    \caption{(Color Online) Same as Fig.~\ref{sq_0.00} but for $\theta =0.64\pi$ in dual $\mathbb{Z}_2$-vortex crystal phase.}
    \label{sq_0.64}
  \end{center}
\end{figure}

Around $\theta_0$, we find a dual $\mathbb{Z}_2$-vortex crystal phase ($0.6\pi<\theta<0.7\pi$), which is the dual phase of $\mathbb{Z}_2$ VC phase, as shown in Fig.~\ref{phase}.
The structure factors at $\theta=0.64\pi$ are shown in Fig.~\ref{sq_0.64}. 
As in the case of the $\mathbb{Z}_2$-vortex crystal phase (see Fig.~\ref{sq_0.16}), $S^{yy}(\bvec{q})$ and $S^{zz}(\bvec{q})$ are asymmetric with respect to the $\Gamma$-M line but symmetric in $S^{xx}(\bvec{q})$.

\begin{figure}[t]
  \begin{center}
    \begin{tabular}{r}
      \begin{minipage}{0.5\hsize}
        \begin{center}
          \includegraphics[clip, width=10pc]{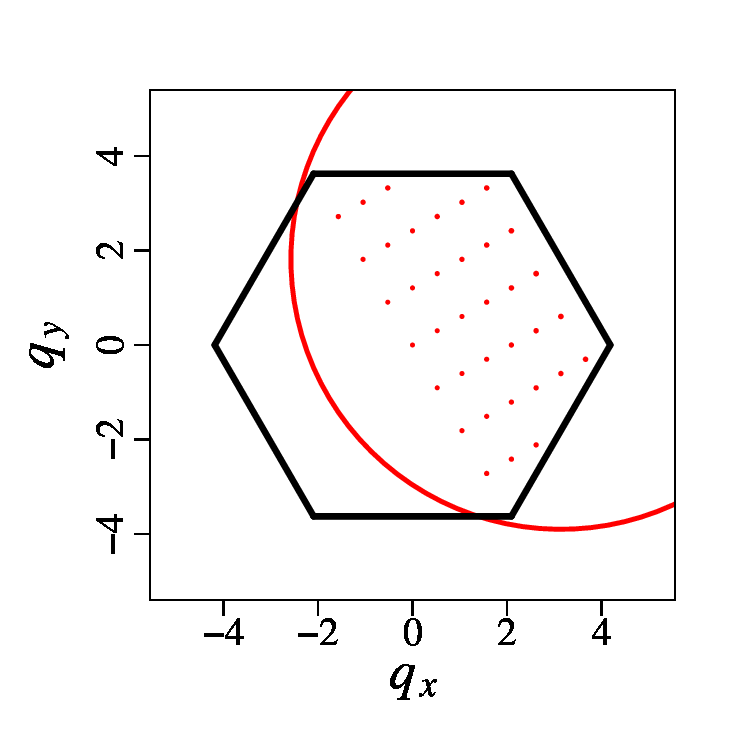}
          \hspace{1cm} (a)
          \vspace{0cm}
        \end{center}
      \end{minipage}
      \begin{minipage}{0.5\hsize}
        \begin{center}
          \includegraphics[clip, width=10pc]{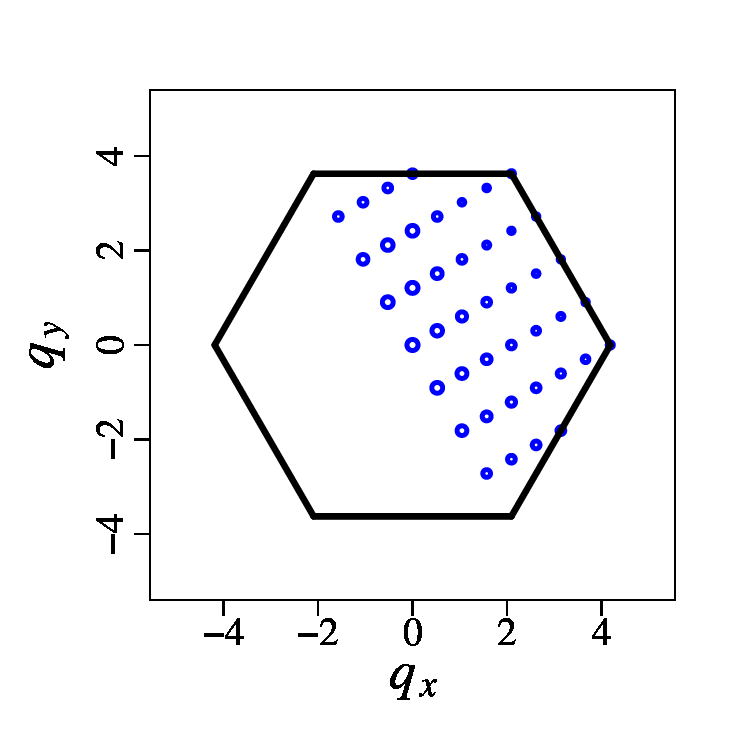}
          \hspace{1cm} (b)
          \vspace{0cm}
        \end{center}
      \end{minipage}
      \\
      \begin{minipage}{0.5\hsize}
        \begin{center}
          \includegraphics[clip, width=10pc]{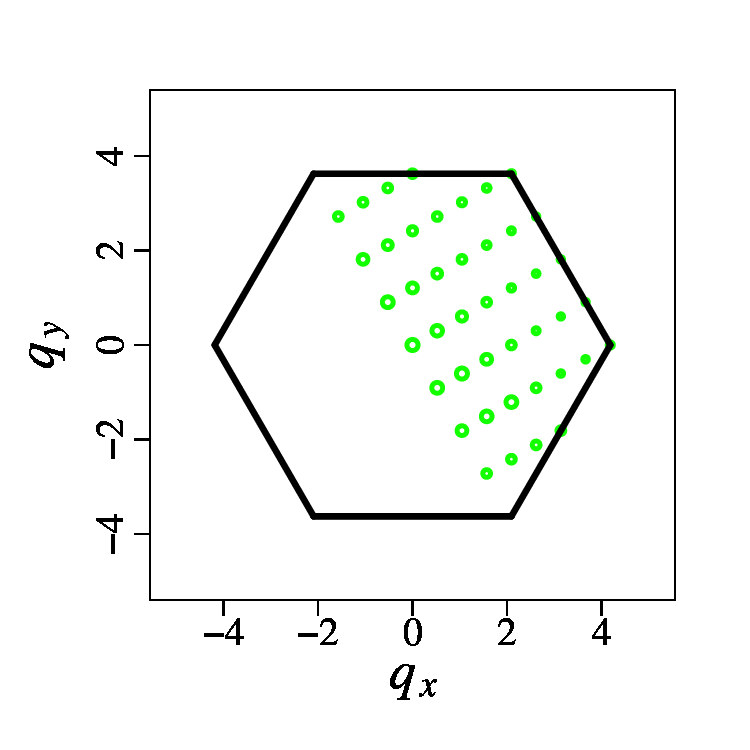}
          \hspace{1cm} (c)
          \vspace{0cm}
        \end{center}
      \end{minipage}
      \begin{minipage}{0.5\hsize}
        \begin{center}
          \includegraphics[clip, width=10pc]{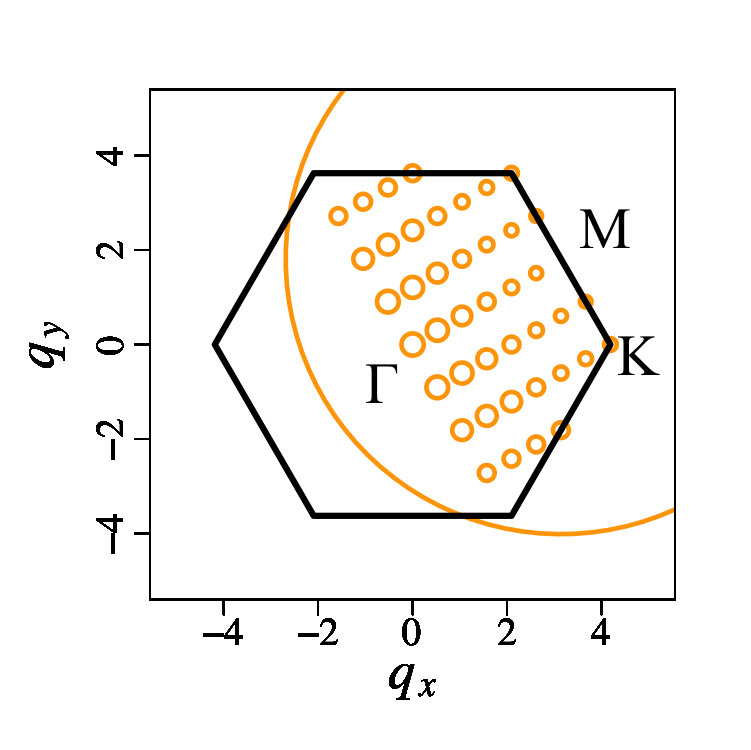}
          \hspace{1cm} (d)
          \vspace{0cm}
        \end{center}
      \end{minipage}
    \end{tabular}
    \caption{(Color Online) Same as Fig.~\ref{sq_0.00} but for $\theta =1.58\pi$ in dual $\mathbb{Z}_6$ FM phase. The center of the large circle in (a) and (d) is located at the M point. }
    \label{sq_1.58}
  \end{center}
\end{figure}

\begin{figure}[htb]
  \begin{center}
    \begin{tabular}{r}
      \begin{minipage}{0.5\hsize}
        \begin{center}
          \includegraphics[clip, width=10pc]{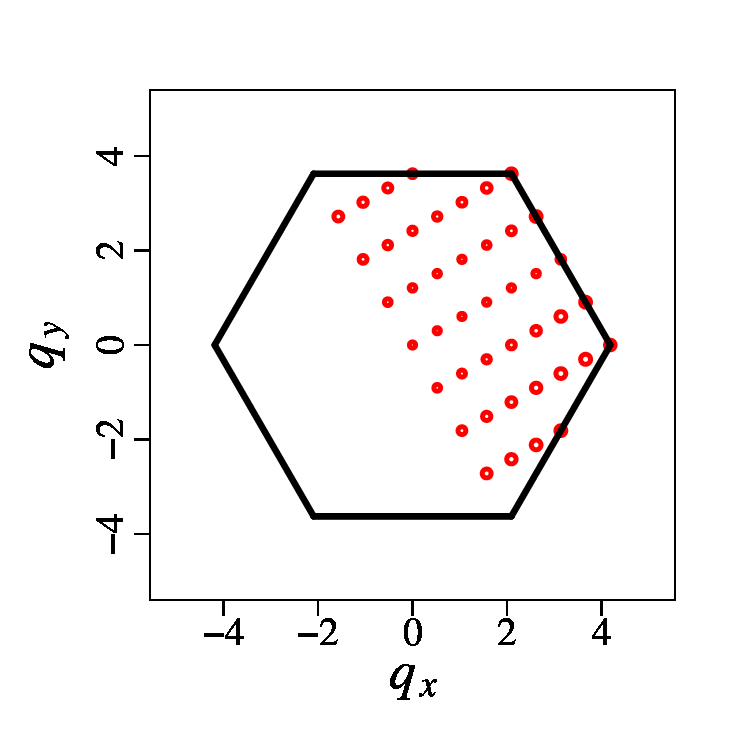}
          \hspace{1cm} (a)
          \vspace{0cm}
        \end{center}
      \end{minipage}
      \begin{minipage}{0.5\hsize}
        \begin{center}
          \includegraphics[clip, width=10pc]{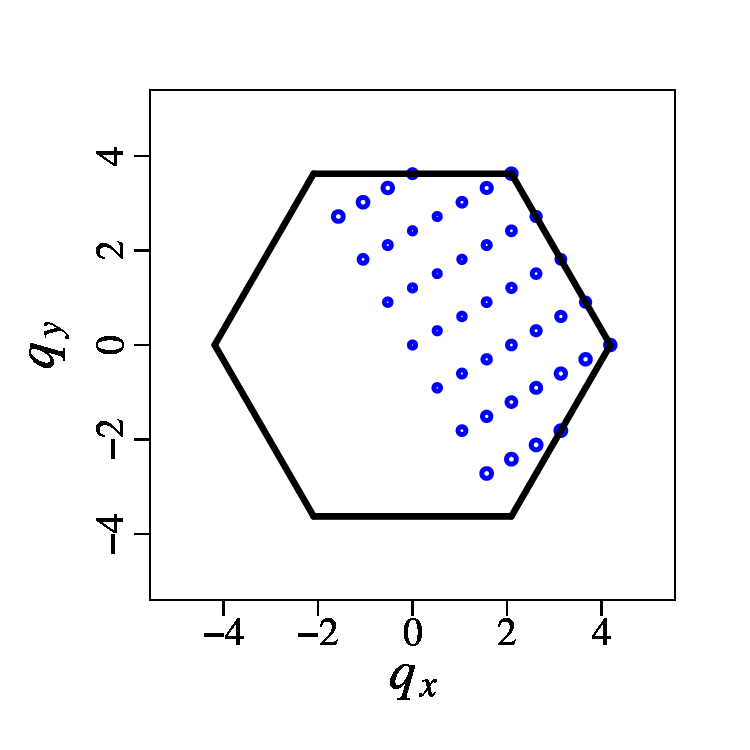}
          \hspace{1cm} (b)
          \vspace{0cm}
        \end{center}
      \end{minipage}
      \\
      \begin{minipage}{0.5\hsize}
        \begin{center}
          \includegraphics[clip, width=10pc]{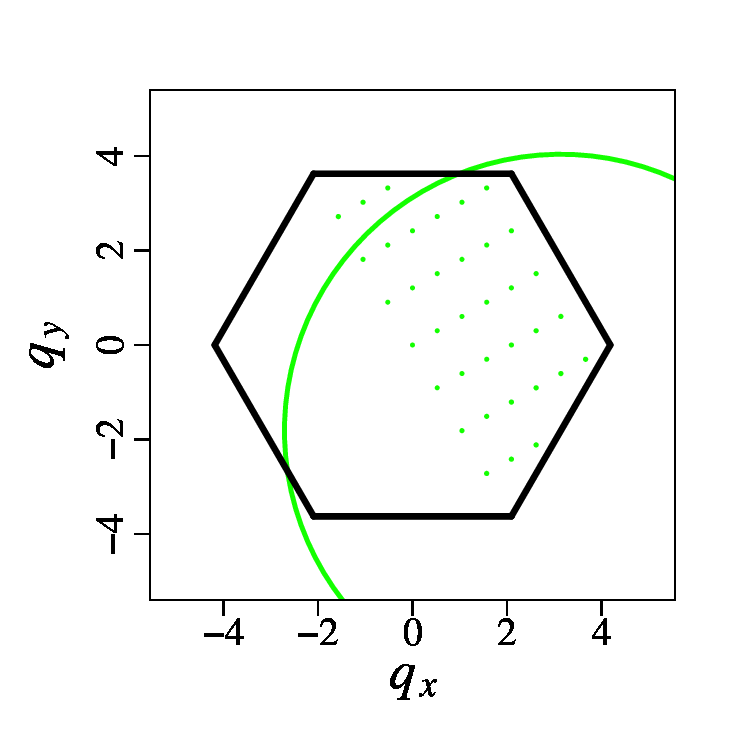}
          \hspace{1cm} (c)
          \vspace{0cm}
        \end{center}
      \end{minipage}
      \begin{minipage}{0.5\hsize}
        \begin{center}
          \includegraphics[clip, width=10pc]{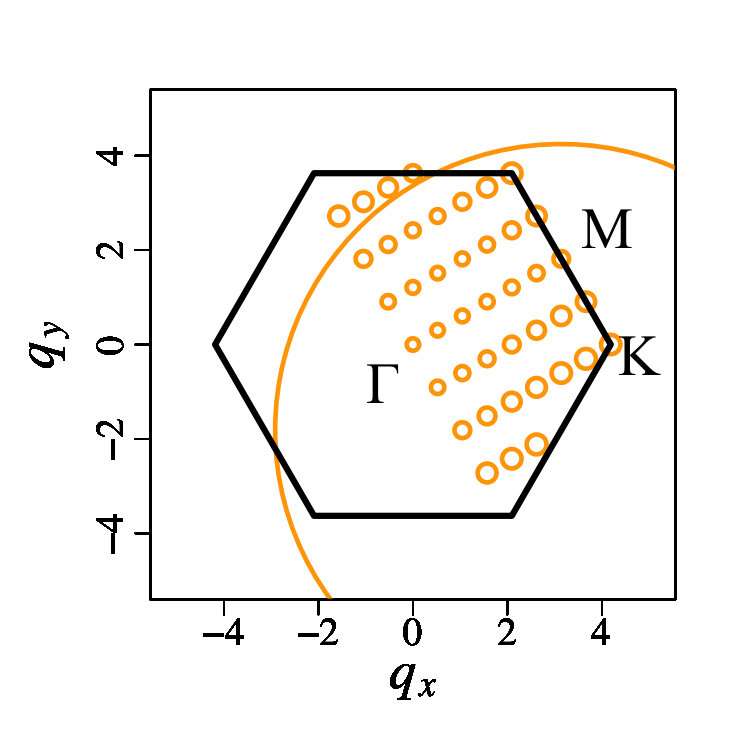}
          \hspace{1cm} (d)
          \vspace{0cm}
        \end{center}
      \end{minipage}
    \end{tabular}
    \caption{(Color Online) Same as Fig.~\ref{sq_0.00} but for $\theta =1.70\pi$ in dual $\mathbb{Z}_6$ FM phase. The center of the large circle in (a) and (d) is located at the M point with negative $q_y$. }
    \label{sq_1.70}
  \end{center}
\end{figure}

The dual FM phase, corresponding to the $\mathbb{Z}_6$ FM phase at $0.68\pi<\theta<1.50\pi$, is located at $1.50\pi<\theta<1.88\pi$.
Figures~\ref{sq_1.58} and \ref{sq_1.70} exhibit the spin structure factors at $\theta =1.58\pi$ and $\theta =1.70\pi$, respectively. $S^{xx}(\bvec{q})$ becomes a stripe-type phase with the largest intensity at the M point with $q_y>0$ in Fig.~\ref{sq_1.58}, while $S^{zz}(\bvec{q})$ becomes largest at another M point with $q_y<0$ in Fig.~\ref{sq_1.70}.
The difference between the spin component of the two cases can be understood by considering spin rotations~\cite{becker2015}. When the $x$ ($z$) component of the spin is ferromagnetically ordered, a dual phase becomes a stripe-type AFM phase where the $x$ ($z$) component is ordered.
However, the directions of these stripes, i.e., the directions of the FM chains, are different from each other by $\pi/3$.

\subsection{Entanglement entropy and entanglement spectrum}
\label{sec3.3}

The entanglement of a wave function can provide useful information on quantum states.
It is measured by the entanglement entropy and entanglement spectrum~\cite{li2008}.
In a system composed of two subsystems A and B, the Schmidt decomposition of a many-body state $|\psi \rangle$ reads
\begin{equation}
|\psi \rangle = \sum _i p_i |\psi^i_A \rangle |\psi ^i _B \rangle
                   =\sum _i e^{-\xi_i} |\psi _A^i \rangle |\psi _B^i \rangle ,
\end{equation}
where $p_i$ is an eigenvalue of the reduced density matrix $\rho _A ={\rm Tr}_B |\psi \rangle \langle \psi | =e^{-\mathcal{H}_E}$ for subsystem A (or $\rho _B ={\rm Tr}_A |\psi \rangle \langle \psi |$ for subsystem B).
The distribution of $\xi_i$ is called the entanglement spectrum, where $\xi_i$ is an eigenvalue of the entanglement Hamiltonian $\mathcal{H}_E$.
The von Neumann entanglement entropy containing non-local topological properties can be written as 
\begin{equation}
S_\mathrm{E} =-\sum _i p_i \ln p_i = \sum _i \xi _i e^{-\xi _i}.
\end{equation}
As subsystem A, we take half of the whole system throughout this paper.
When calculating the entanglement entropy, we consider a system with cylindrical geometry due to the cylindrical boundary conditions since the boundary of the system and environment blocks is clear and easy to define.
We calculate the reduced density matrix of the subsystem indicated by the orange shaded area in Fig.~\ref{triangle}.
As we have seen in the previous section, there is no Kitaev spin-liquid phase in the KH model on a triangular lattice, and therefore no degeneracy appears in the entanglement spectrum, in contrast with the case of a honeycomb lattice\cite{shinjo2015}.

In one-dimensional systems, the following scaling behavior of the entanglement entropy is well understood: in the thermodynamic limit, the entropy is finite for gapped systems, while it is logarithmically divergent for gapless systems~\cite{holzhey1994, calabrese2004}.
The scaling behavior is universal, depending only on the central charge and not on the microscopic detail of the systems.
In spite of extensive studies, however, entanglement structures in more than one dimension have not necessarily been resolved.
Although the central charges are not defined, universal terms have also been suggested for even spatial dimensions.
The scaling behavior has been studied and the ground state of a local Hamiltonian is generally believed to produce an ``area law" scaling: the entanglement entropy is proportional to the area of the boundary between subsystems.
A deviation from the area law indicates the existence of certain long-range or non-local correlations.
There are subleading corrections to the scaling behavior that are regarded as universal quantities identifying and characterizing quantum phases and phase transitions.
A well-known example is the topological entanglement entropy of gapped topological systems.
The universal scaling of the entanglement entropy at two-dimensional conformal (scale-invariant) quantum critical points has been studied~\cite{hsu2009, stephan2009, zaletel2011, stephan2011, stephan2013}.
In two-dimensional gapless systems with a broken spontaneous continuous symmetry, subleading corrections can still have universal properties, such as subleading logarithmic correction in two-dimensional systems with Nambu-Goldstone modes~\cite{tagliacozzo2009, metlitski2009, ju2012}.
Therefore, studying the entanglement entropy in various two-dimensional systems is helpful for understanding its behavior in two dimensions.

Figure~\ref{phase}(c) shows the $\theta$ dependence of entanglement entropy.
At phase transition points, the entanglement entropy discontinuously changes.
The slight deviation of the discontinuity from $\theta=1.5\pi$ is due to finite-size and/or boundary effects since a discontinuity is observed at $\theta=1.5\pi$ for the toroidal boundary conditions (not shown).
The discontinuous change is simply due to the fact that all the phase transitions appearing in the KH model on a triangular lattice are of first order: at the first-order phase transition point, the ground state suddenly changes to another state with different entanglement.

\begin{figure}[t]
          \vspace{0.5cm}
\begin{center}
\includegraphics[width=18pc]{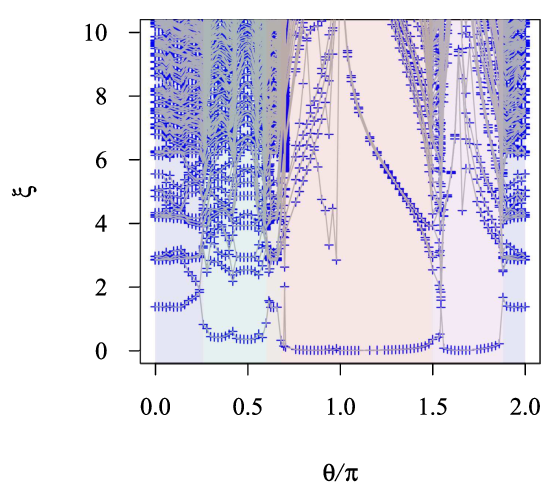}
\caption{(Color Online) Entanglement spectrum for the KH model in Eq.~(\ref{KHmodel}) on a triangular lattice. Blue pluses represent entanglement levels and black lines connect the spectra belonging to the same entanglement levels.}
 \label{es}
 \end{center}
\end{figure}

The entanglement spectrum $\xi_i$ for the KH model is shown in Fig.~\ref{es}, where low-lying entanglement levels are plotted from the smallest value starting from $i=1$.
The spectral distribution of the entanglement spectrum changes with the parameter $\theta$.
In the case of the KH model on a honeycomb lattice, the degeneracy of the entanglement spectrum has been found in a Kitaev spin-liquid phase, which is due to the intrinsic nature of the Kitaev spin liquid~\cite{yao2010, shinjo2015}.
On the triangular lattice, however, there is no such spin-liquid phase, and thus the degeneracy of the entanglement spectrum is not found except at phase transition points.
To characterize phase transition points using the entanglement spectrum, we consider the Schmidt gap $\xi _2 - \xi _1$,~\cite{dechiara2012} where $\xi _1$ and $\xi _2$ are the lowest and second-lowest levels of the entanglement spectrum, respectively.
In contrast to the KH model on a honeycomb lattice, the Schmidt gap in each phase is clearly larger than other gaps at higher levels. This implies that the lowest entanglement level $\xi _1$ is well separated from the other levels and thus predominately contributes to the ground-state wave function of the total system.
The Schmidt gap closes at phase boundaries, where the eigenstates of the entanglement spectrum cross and interchange with each other.
The crossing is seen only near the boundaries since the ground state changes to other states with a first-order phase transition.
Therefore, the Schmidt gap shows a singularity at phase transition points in the KH model on the triangular lattice.

In addition to the phase boundaries, the entanglement entropy and entanglement spectrum can detect changes in the symmetry of wave functions. For example, there is a peak structure of the entanglement entropy at $\theta=0.4\pi$ in the nematic phase, whose dual point at $\theta=0.6\pi$ also shows a peak structure. This is an indication of a change in the symmetry of the ground-state wave function without phase transition. Interestingly, there is no noticeable change in the spin structure factor at these values of $\theta$. Detecting such a change in the wave function is important since the symmetry of the ground-state wave function is not necessarily determined by the symmetry of the KH model related to the Klein duality.

\section{Summary and Outlook}
\label{sec4}

We have studied the KH model on a triangular lattice by using the DMRG method and constructed a full-parameter ground-state phase diagram.
We have used a large cluster with $12\times 6$ sites and found a 120$^\circ$ AFM phase, a $\mathbb{Z}_2$-vortex crystal phase, the dual $\mathbb{Z}_2$-vortex crystal phase, a nematic phase, a $\mathbb{Z}_6$ FM phase, and its dual, in good quantitative agreement with previous studies~\cite{rousochatzakis2012, becker2015}.
The magnetic structures of these phases have been determined by calculating the spin structure factors for each spin component.
As the spin structure factors suddenly change at all phase boundaries, we conclude that all the phase transitions are of first order.

In the $\mathbb{Z}_6$ FM phase and its dual FM phase, we have found that the spin structure factors change at the SU(2) symmetric points $\theta =\pi$ and $\theta = 1.64\pi$, respectively. 
There are no phase transitions but the direction of ordering changes.
Since the $\mathbb{Z}_6$ FM phase is generally 6-fold degenerate, one of them will be selected by external conditions. In the geometry of our $12\times 6$-site lattice, the FM orderings in different directions are selected across SU(2) points.

In the $\mathbb{Z}_2$-vortex crystal phase, the spin structure factors of the $y$ and $z$ spin components become asymmetric with respect to the $\Gamma$-M line, as expected from the previous studies~\cite{rousochatzakis2012, becker2015}.

The entanglement entropy markedly changes at all phase boundaries, as is expected from the first-order nature of the transitions.
This is in contrast to the extended KH model on a honeycomb lattice studied previously~\cite{shinjo2015}, where a topological spin-liquid phase exists.
By examining the entanglement spectrum, we have found that the Schmidt gap is much larger than other entanglement gaps.
This implies that the lowest entanglement level is crucial for describing the ground-state wave function of the whole system.
At phase transition points, the Schmidt gap closes, and therefore the change in the entanglement structure is clear in the KH model on the triangular lattice.

We have confirmed the exotic magnetic phases proposed in the previous studies of the triangular KH model~\cite{rousochatzakis2012,becker2015}.
It is thus interesting to investigate the presence of such phases in the candidate spin-liquid material Ba$_3$IrTi$_2$O$_9$~\cite{catuneanu2015}.
In particular, our results for spin structure factors $S^{\gamma \gamma}(\bvec{q})$ could be useful for analyzing polarized neutron scattering data obtained in the near future.

We have found no quantum spin-liquid phase in the KH model on the triangular lattice, in contrast to the KH model on the honeycomb lattice.
However, a quantum spin liquid might be stabilized by additional interactions such as further-neighbor interactions.
In fact, a previous study proposed a chiral spin-liquid phase close to the AFM Kitaev limit~\cite{li2015}.
Furthermore, Ba$_3$IrTi$_2$O$_9$ has been suggested to have a spin-liquid state~\cite{dey2012}.
Therefore, investigating a quantum spin-liquid state in an extended version of the KH model will be interesting.

\begin{acknowledgments}
K.S., S.S., and T.T. acknowledge Y. Yamaji, T. Okubo, N. Kawashima, and M. Imada for useful and stimulating discussions. K.S. acknowledges the support from RIKEN as a Junior Research Associate (JRA).
This work was financially supported by MEXT HPCI Strategic Programs for Innovative Research (SPIRE) (hp140215, hp140136, hp150211) and the Computational Materials Science Initiative (CMSI). The numerical calculation was partly carried out at the K Computer, RIKEN Advanced Institute for Computational Science, and the Supercomputer Center, Institute for Solid State Physics, University of Tokyo. This work was also supported by Grants-in-Aid for Scientific Research (Nos. 25287096 and 26287079) from MEXT, Japan.
\end{acknowledgments}

\end{document}